\documentclass[a4paper,11pt]{article}

\usepackage{amsmath,amsfonts,amssymb,times,mathptmx,graphicx,amsthm}
\usepackage{placeins}
\usepackage[T1]{fontenc}
\usepackage[latin1]{inputenc}

\title{\bf Properties of the Center of Gravity as an Algorithm for
Position Measurements }

\author{Gregorio Landi\\
Dipartimento di Fisica, Universita' di Firenze, Italy\\
 and INFN Sezione di Firenze, Firenze,Italy\\   \\
{\em  April 14, 2001 and August 13, 2019 }}

\date{ }
\begin{document}
\maketitle
\pagenumbering{arabic}
\begin{abstract}
The center of gravity \(x_{g}= \sum_{i}E_{i} x_{i}/\sum_{i}
E_{i}\) as an algorithm for position measurements is carefully
analyzed. Many mathematical consequences of discretization are
extracted. The origin of the systematic error of the algorithm is
shown to be connected to the absence of  band limits in the
Fourier Transform of the signal distributions, which, owing to the
intrinsic properties of the measuring devices, must have a finite
supports. However, special signal distributions exist among the
finite support functions which are free from the discretized
error. In the presence of crosstalk, it is proved that some
crosstalk spreads are able to eliminate the discretization error
for any shape ({\em ideal detector}).  For all other cases, analytical expressions and
prescriptions  are given to correct the error and to efficiently
simulate various experimental situations.

\end{abstract}
\tableofcontents

\section{Introduction 2019}\indent

The aim of this paper (published in {\em Nuclear Instruments and Methods $\cdots$}  A 485 2002 698-719) 
was the discussion and the calculation of the systematic error of
the center of gravity (COG) (sometimes called weighted average or barycenter) as a
positioning algorithm in one-dimensional geometry. The two-dimensional geometry
is discussed in another paper. The arguments described here are relevant for recent 
developments.

The existence of the COG systematic error was known by a long time. In fact, one of our first work in
high energy physics was just the elimination of the COG systematic error in the 
electron reconstruction of the Crystal Ball experiment. We never published the method, 
it used higher-order tensors and was very fast and precise. It was that experience
that raised our interest in the problem, but only many years later we found the right mathematical
tools (the Shannon sampling theorem~\cite{11}) to handle the problem. 
Our numerical simulations produced plots very similar to diffraction 
patterns in optics (as fig.2), but no waves are contained in the COG 
expressions. Where are the waves? The properties of the Fourier transform
contained in the Shannon sampling theorem were the sought waves. This is the
mathematical reason of our extensive (excessive?) use of Fourier 
transform and Fourier series. In any case the Fourier transforms are simple
objects compared to the Bessel functions and the machinery required to extend 
the method to a sphere. We did it, but even the hard-disk considered it too 
complicated and decided to crash with a loss of all the work. We never did it again,
but it is a possible task.        

An important step was the introduction of the two-strip COG
correction in 1983: the $\eta$-algorithm~\cite{16}. However, analytical
expressions of this error (if any) was lacking, and a consistent elimination of it
requires analytical form (or forms). The importance of the elimination of any systematic error
in a fit is accurately discussed by Gauss in his paper of 1823. Here, Gauss stated that it is incorrect 
to handle the systematic errors as random errors after the availability of a "table" for 
those errors. It is natural to suppose that analytical expressions of
systematic errors, as those of this paper, are equivalent (or better) than the tabular
forms indicated by Gauss. 
Thus, applying the Gauss criterion, each paper
that uses the COG positioning in a fit, without the proper correction of the systematic
error, is incorrect at least from the date (2002) of this paper.
It is evident that many papers were published with this type of error in the last
years. This looks as a drop of the physics awareness that, without referring to the Gauss
paper, knew very well the danger of the systematic errors.

Stimulated by these uncorrect papers, but independently from the Gauss paper,
we proved the dangerous effects of the neglect
of the COG corrections in ref.{\em arXiv:1606.03051}. That paper was a first version
dedicated to the momentum reconstructions (the final version is published in
{\em INSTRUMENTS 2018 2 22} and {\em arXiv:1806:07874}).
In one of the tests, the random noise was almost completely eliminated in the simulations.
The distributions of the momentum reconstructions, based on positioning corrected by
the COG systematic errors (eq.25 for example),
showed a rapid convergence toward Dirac $\delta$-functions as expected. On the contrary,
the momentum distributions based on the simple COG positioning remained invariant to the
noise suppression: the detector quality (or better the signal-to-noise ration) becomes
irrelevant in large extent.  This result is in a perfect agreement with the Gauss warning
about unpleasant effects given by systematic errors.

However, the study of the COG systematic error gave us a direct evidence of the necessity
of a correction for each hit. A natural consequence was the necessity to construct 
a different probability distribution for each hit. Parts of the mathematical 
developments, discussed here, were used in the paper {\em arXiv:1404:1968}
(published by {\em JINST 2014 9 P10006} ) for the construction of this type of 
probability distributions.

Among the by-products of the new probability distributions, an interesting effect
was illustrated in {\em INSTRUMENTS 2018 2 22}: an approximate linear growth of the
momentum resolution with the number $N$ of detecting layers. This growth is much faster than
the $\sqrt{N}$ of the standard fit. To discuss better this linear growth, a simpler
simulation with straight tracks was reported in ref. {\em arXiv:1808.06708}. Even here 
it is obtained a very large linear increase in the resolution of the fit parameters.
To simplify, we will always speak of linear growth, even if small deviation from 
linearity are present.    
A Gaussian toy-model was introduced with a very simplified form of weights. This model
is an easy illustration of the mechanism producing the linear growth. 
Among the simplified  forms of weights, it was possible to connect 
the hit weights with the COG probabilities: the lucky-model.

The correct implementation of the lucky-model (as it was defined in {\em arXiv:1808.06708}),
requires the elimination of the COG systematic errors or a their drastic reduction. 
The appropriate use of the analytical expressions for the COG can be an 
alternative to the more complicated $\eta$-algorithm.
The lucky-model is a very economical tool for a drastic improvement of the fit
resolution. The anomalies described here for the COG with a small number of strips must be
accurately managed to use the lucky-model at large angles. Interesting enough, the inverse 
of effective weights of the lucky-model are strictly connected with eq.27 that 
describes the average signal distribution collected by a strip.

A line with the definition of the ideal detector is added at the end of subsection 6.6, it was forgotten
in the writing of the paper. The subsection 6.6 describes a special kind of
cross-talk that is able to suppress the COG systematic error for any form of signal distributions.
The COG of this ideal detector is evidently compliant to the Gauss prescriptions.
Detectors with floating strips tend to this ideal conditions, but the COG corrections
are important even in this case.

We corrected few other printing errors, the formatting in a two columns style added few errors
undetected at the time of the proof reading.

\section{Introduction} \indent

The present generation of high-energy physics experiments shows a
dramatic increase in the detector segmentation to capture the
thinnest detail of the particle signals~\cite{atlas,2}. These improvements
must be completed together with  improvements in the
reconstruction algorithms. However, a  reconstruction is always a
pattern recognition, and the well-known  ill-mathematical
definition of any pattern recognition  poses a serious problem to
its generalization. Hence, almost always, the reconstruction
algorithms are largely blended with ad hoc recipes to extract the
best of the detector itself. Even if quite successful, their
improvements are difficult to separate from the details of the
detector, leaving few possibilities of exportation.

The aim of this work is to analytically treat the center of
gravity (COG) algorithm, with  particular attention to the effects
of discretization. These results are mostly mathematical. For this
reason, they are not bound by the above limitations and are
applicable wherever the assumptions fit in the application. The
use of the COG for position (and other) measurements is extremely
widespread in scientific and practical applications, which are far
too numerous to list here. Actually, only in rare cases can data
analysis avoid calculating the COG of some signal distribution. We
are primarily interested in high-energy physics applications, and
our attention is focused on the topics we know best, as are our
references.

At a superficial glance, widespread use of the algorithm is not
unexpected: It actually looks very easy and easily justifiable. In
fact the COG coincides with a symmetry point (line or plane) of
the system. Hence, if the average image of the measured object on
the measuring device has a symmetry property (point, line, or
plane), the COG could give an estimation of its position. One
drawback of this justification resides in the discretization that
any automatic measuring device must perform on the image. Almost
always, the discretization destroys any symmetry on the system,
and the use of the COG introduces a systematic error in
measurement which will be called even  discretization error due to
its origin in the discretization of the signal collection.

In high-resolution measurements, the  presence of a systematic
error in the algorithm is well known, and many empirical
procedures have been developed to reduce its effect. Often, this
fine-tuning requires lengthy and expensive Monte Carlo simulations
or complex integration, with considerable probability of losing
control of important  details in the problem. More versatile and
transparent methods are evidently of great help. No specific
instrument will be considered, even if the declared polarization
toward  high-energy physics applications will be evident.

Here, we will limit ourselves to unidimensional geometry;  the
extension to  bidimensional geometry is straightforward in some
major cases.  We will find  classes of functions whose COGs are
free of discretization error, but these  results are probably of
scarce use, since the selection of the signal distribution is
rarely possible. An interesting consequence of these special
functions  is found on the crosstalk function. It turns out that
if the crosstalk function is any of the above functions, the
crosstalk saves the "energy" of the signal. Special crosstalk
functions whose COGs are free of discretization error for any
signal distribution will be defined.

We will determine analytical relations of continuous (almost
everywhere) functions and their histograms. These relations are
central to our study of the discretization effects. In its
simplest form, the signals collected by a set of detectors are the
histograms of the continuous signal distribution, and the bin size
is equal to the size of the elementary detector (pixel, strips,
crystal, etc.).

In Section 2, we introduce the problem with all the required
definitions and examples. The systematic error is calculated with
the easiest assumptions. We shall recall elements of signal
theory, limiting ourselves to the strict necessities of the
problem. All the derivations are based on standard properties of
the Fourier Transforms (FT) and the Fourier Series (FS).

In Section 3, the conditions of the signal distributions which
grant the absence of  systematic error are discussed. Only
exceptionally will the experimental signal distributions pertain
to these classes and be free from error:  For all other cases,
equations will be given to relate true position with the COG.

A first analytical relation of a signal distribution with the
histogram will be given in Section~4. This equation allows a
direct simulation of the signal collection in the measuring
device; it can easily be completed with noise models for the
fine-tuning of empirical relations which go beyond the COG. We
will define and apply a method for summing the first part of the
double series encountered in the equations.

In Section 5, we extend this approach to the case of signal loss
and crosstalk, defining their analytical aspects and properties.
This allows a detailed treatment of a few reconstruction methods
and experimental setups. The effects of  crosstalk on signal
collection and COG reconstruction are discussed in detail along
with some of the unexpected  properties of the crosstalk. Methods
are developed to handle the suppression of the low-signal sensors.
The discontinuities generated by these cutoffs are isolated.
Reconstruction of  position from a distribution of COG
measurements is discussed and a few of its consequences and
defects are pointed out.

Section 6 will address the study of the noise and signal
fluctuations. A method for extracting fluctuations from a Monte
Carlo simulation will be illustrated.

\section{ Sampling methods}
\subsection{ Schematic properties of detectors} \indent

To accurately describe our problem, we shall consider the signal
distribution of a high-energy photon showering in a homogeneous
crystal detector with the geometry of the BGO  calorimeter of the
L3 experiment ~\cite{3} or the \(\mathrm{PbWO}_{4}\) calorimeter of the CMS
experiment~\cite{4}.  Roughly speaking, these detectors consist of an
array of truncated-pyramid shape crystals whose longest axis
points toward the interaction center (if we neglect the small
off-center distortion needed to reduce the energy loss at the
crystal borders). Photons coming from the interaction  point
produce showers in an almost homogeneous medium, and their average
has an axially symmetric shape whose symmetry axis coincides with
the photon direction. The light collection  projects the shower on
a plane orthogonal to the photon direction and the symmetry point
of the  average distribution is the impact position of the photon.
The readout system generates a set of numbers which in
well-designed calorimeters  are proportional to the energy
released by the photon shower in each crystal~\cite{5}. Each detecting
element can be approximated to a square pixel identical to all the
others, and arranged in a regular flat orthogonal array. This
bidimensional distribution can often be explored as two orthogonal
unidimensional distributions due to the linearity of COG in the
energy distribution. It is evident that this schematization is
useless in a complex geometry such as that of the Crystal Ball
detector~\cite{5}.

An almost perfect unidimensional system is given by silicon micro
strip detectors in a tracker system~\cite{6,7}. Here, a particle
traversing the detector  distributes signals on a few contiguous,
very long (infinite for our needs) parallel strips. The
three-dimensional initial ionization has an average symmetry axis
directed along the path of the incoming particle.  The detector
geometry allows the retrieval of discrete information on a
unidimensional signal distribution   $\psi(x)$.  Here, and in the
following, x is a reference axis on the detector plane,
perpendicular to the strip direction, with its origin in the
middle of a strip. For particle directions orthogonal to the strip
plane, the point of symmetry of the average on $\psi (x)$ is the
position of the impact point.

\subsection{Average signal distribution} \indent

By $\varphi(x)$, we will indicate the average signal distribution
for the impact point  $\varepsilon=0$. The function $\varphi(x)$
and its discrete reduction performed by the readout system will be
the focus of the present investigation.  Let us first consider the
properties of  $\varphi(x)$ which will be required in the
following:

a) $\varphi(x)$ is real and symmetric around zero   $\varphi(x)=
\varphi(-x)$

b)  $\varphi(x)$ is continuous almost everywhere and has a
continuous and derivable FT

c) $ \varphi(x-\varepsilon )$ is the signal distribution in a
homogeneous medium when the impact point is $\varepsilon $.

We will call $\varepsilon$  the impact point, in the jargon of
high-energy physics. In general, $\varepsilon$ is the position of
the symmetry point of the average signal distribution for any kind
of signal source. Our results are valid even for asymmetrical
$\varphi(x)$. In this case, $\varepsilon$ is the position of the
COG of $\varphi (x)$, but in the equations we will explicitly use
the symmetry of $\varphi(x)$.

To simplify the notation, we normalize $\varphi(x)$ as
$\int_{-\infty} ^{+\infty} \varphi (x) dx=1$.  In addition, in
some derivations, we will need $\varphi(x)$ to be positive with a
single maximum.

From points a) and b), it is clear that $x_{g}$ defined as
$x_{g}=\int_{-\infty} ^{+\infty} x\varphi (x) dx$ to be the COG of
$\varphi(x)$ is zero. The impact point $x_{i} = \varepsilon$,
yields the identity $x_{g}=\varepsilon$.  All the values of
$x_{g}$ are independent of the details of signal distribution and,
at this level, \textit{the method is a perfect position measuring
device}. However, $\varphi(x)$ is not accessible by the
experiments nor it is normalized. The readout system performs  a
discrete reduction that drastically modifies the method's
measuring effectiveness. In its simplest form, the discrete
reduction amounts to a set of finite disjointed integrations on
the signal distribution, spoiling it from any simple connection to
$\varphi(x)$ and to $\varepsilon$. So, the readout data consist
of a set of numbers (very few indeed) $\alpha_{n}(\varepsilon)$
defined by:
\[\alpha_{n}(\varepsilon)=\int_{n\tau-\tau/2}^{n\tau+\tau/2}
\varphi(x-\varepsilon)dx \]
where $\tau$ is the distance of the strip axis (or the axis of any
other detector array) and  $\alpha_{n}(\varepsilon )$ is the
signal collected by the strip $n$ (or the sum of the signals
collected by a line of detectors orthogonal to the x-axis). Now,
the COG is defined by $x_{g}=\sum_{n}\alpha_{n}(\varepsilon)\tau
n/\sum_{n}\alpha_{n}(\varepsilon )$. The reference system has its
zero on the middle of the detector with the maximum signal, and we
can limit ourselves to exploring the region with
$|\varepsilon|\leq1/2$. Excluding the case of $\tau\rightarrow 0$
and $n\rightarrow \infty$, it is evident that the discretized
$x_{g}$ differs from $\varepsilon$ almost everywhere.

\subsection{Sampling} \indent

To obtain explicit  relations, we shall use a few elements of
signal theory. As is often done in signal theory,  we can rescale
all the lengths to have $\tau=1$ without loss of generality, but
the practical use of the equations can entail some ambiguities
which might well be avoided. The tradeoff is an additional symbol
to handle, but, at the same time, the equations retain the proper
dimensions. First, we have to define the functions:
\begin{align}
&f(x)=\int_{-\infty}^{+\infty}\Pi(\frac{x-x'}{\tau})
\varphi(x')\mathrm{d}x' \ \nonumber \\
&f_{\varepsilon}(x)=\int_{-\infty}^{+\infty}\Pi(\frac{x-x'}{\tau})
\varphi(x'-\varepsilon)\mathrm{d}x',
\end{align}
where  $\Pi(t)$ is the interval function $\Pi (t)=1$ for $|t|<1/2$
and $\Pi(t)=0$ elsewhere, and wherever needed, $\Pi(\pm1/2)=1/2$.
This is one of the simplest linear transforms. More complex cases
will be explored in the following sections. Due to the symmetry of
$\Pi(t)$ and  $\varphi(x)$, function $f_{\varepsilon} (x)$ is
symmetric with respect to $ x=\varepsilon$ . The properties of
convolutions~\cite{8} state that $f_{\varepsilon} (x)$  is equal to $f
(x-\varepsilon )$ and the COG of  $f_{\varepsilon}  (x)$ is the
sum of the COGs of  $\Pi (x)$ and $\varphi(x-\varepsilon )$. The
first is zero, the second is $\varepsilon$ , and the COG of
$f_{\varepsilon} (x)$ is that of $\varphi(x- \varepsilon)$. Now,
\begin{figure}[h!]
\begin{center}
\includegraphics[scale=0.90]{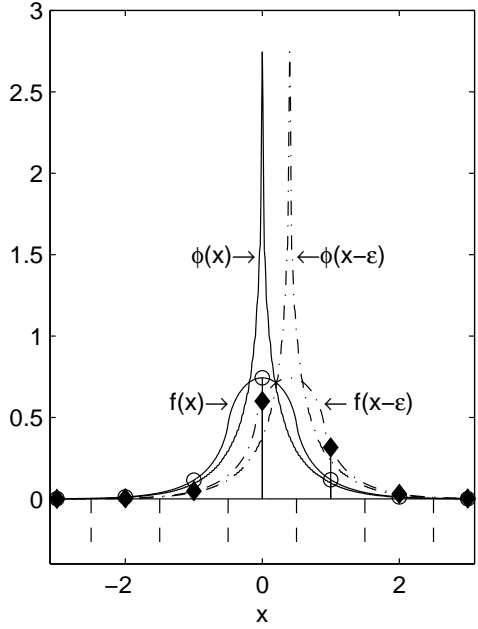}
\end{center}
\caption{\em $\phi(x)$ is the signal distribution with zero impact
point, {\rm f}(x) is its convolution with an interval function.
The dots are the signals collected by the detectors centered at
x=-3,-2,-1,0,1,2,3 and are samples of {\rm f}(x).
$\phi(x-\varepsilon)$, {\rm f}$(x-\varepsilon)$ and the diamonds
are as above for an impact point $\varepsilon$. Dashed lines are
the borders of the detectors. }
\end{figure}
the $\alpha_{n}(\varepsilon )$ values are given by the sampling of
$f_{\varepsilon}  (x)$ at each $n\tau$ $(n=0,\pm1,\pm2,....)$ and
they are the signals collected by the detectors whose axes is
located at points $x=n\tau$ as shown in Figure 1. The set
$\{\alpha_{n}(\varepsilon )\}$ can be formally expressed as a
function of $x$ with a series of  Dirac $\delta$-functions:
\begin{align}
&s_{\varepsilon}(x)=\sum^{+\infty}_{n=-\infty}\alpha_{n}(\varepsilon)\delta(x-n\tau)
\\
&s_{\varepsilon}(x)=f(x-\varepsilon)\sum^{+\infty}_{n=-\infty}\delta(x-n\tau).\nonumber\tag{2'}
\end{align}
%
Defining  $S_{\varepsilon}(\omega))$ the FT of
$s_{\varepsilon}(x)$, $x_{g}$ is given by the relation~\cite{8}:
\begin{equation}
x_g=i\frac{1}{S_{\varepsilon}(0)}\frac{dS_{\varepsilon}(\omega)}{d\omega}\mid_{\omega=0}
\end{equation}
and, defining $F_{\varepsilon} (\omega )$ as the FT of
$f_{\varepsilon}(x)$, the Poisson identity~\cite{9}  relates
$S_{\varepsilon} (\omega )$ with $F_{\varepsilon} (\omega )$:
\begin{equation}
S_{\varepsilon}(\omega)=\tau^{-1}\sum_{k=-\infty}^{+\infty}F_{\varepsilon}(\omega-2\pi
k/\tau).
\end{equation}
%
A sketchy justification of the Poisson identity can be given with
the observation that Eq. (2') is the product of two functions, but
one is periodic with period $\tau$. Its FT is a sum of  Dirac
$\delta$-functions with arguments $(\omega-2\pi k/\tau )$, and the
convolution of $F_{\varepsilon} (\omega )$ with this sum of  Dirac
$\delta$-functions, assumes the form of Eq. (4). According to the
convolution theorem and the shift property of FT,
$F_{\varepsilon}(\omega )$ is equal to:
\begin{equation}
F_{\varepsilon}(\omega)=2\frac{\sin(\omega\tau/2)}{\omega}\Phi(\omega)e^{-i\varepsilon\omega},
\end{equation}
where the first term is the FT of $\Pi(x/\tau)$, and
$\Phi(\omega)$ is the FT of $\varphi(x)$. The function
$\Phi(\omega )$ is real and symmetric, with $\Phi(0)=1$ for the
normalization, and its first derivative is zero at $\omega=0$. The
absence of loss in the signal collection fixes the normalization
of $ s_{\varepsilon} (x)$ as that of $\varphi (x)$; actually, it
is $F_{\varepsilon} (-2\pi k/\tau)=0$ for $k\neq 0$ and $
F_{\varepsilon} (0)=\tau$ giving $S_{\varepsilon} (0)=1$ in Eq.
(4). From Eqs. (3)-(5), and with the properties of $\Phi(\omega)$,
the explicit form of $x_{g}(\varepsilon )$ turns out to be:
\begin{equation}
x_{g}=\varepsilon+\frac{\tau}{\pi}\sum_{k=1}^{\infty}\frac{(-1)^{k}}{k}\sin(2\pi
k\varepsilon/\tau)\Phi(2\pi k/\tau).
\end{equation}
%
The first term of Eq. (6) is  the COG of  $f_{\varepsilon} (x)$ or
of $\varphi (x-\varepsilon )$. It is independent of the shape and
the sampling period and is thus the perfect measuring device we
mentioned. The other terms are the systematic error of the
algorithm due to the sampling of the $f_{\varepsilon} (x)$ at
$\tau$ intervals. Their $\varepsilon$-dependence has the form of
an FS of an odd function of with amplitudes $(-1)^{k}\Phi(2\pi
k/\tau)/k$ and period $\tau$ as conjectured in~\cite{10}. This
periodicity is due to the form of the sampling function in Eq. (2)
which is periodic as well. Physically, this amounts to considering
an infinite row of identical detectors. In Section 5, we will
abandon this condition to handle more realistic cases. It is
evident in Eq.(6) that  $x_{g}=\varepsilon$ for $\varepsilon=0$
and $\varepsilon = \tau/2$ and for all other integer and
semi-integer values of $\tau$, that is a consequence of the
symmetry of our detector setup which survives in these special
points.

If  $\varphi(x)$ converges to a Dirac $\delta$-function, $x_{g}$
reaches the maximum error, as expected. In this case, where
$\Phi(2\pi k/\tau)$ converges to one, the series in Eq. (6) is the
Euler $sine$ series~\cite{8}, and  $x_{g}$ becomes identical to zero
for $|\varepsilon| <\tau/2$ and $x_{g}=\varepsilon$ for
$|\varepsilon |=\tau/2$.

\section{Special Shapes}
\subsection{The absence of discretization error} \indent

Let us now examine the error of $x_{g}$ as expressed by Eq.(6) in
greater detail. The FS form of $x_{g} -\varepsilon$ allows the
analytic calculation of the mean square value on a period
$|\varepsilon |\leq\tau/2$ with the Parseval identity:
\begin{equation}
\Delta^{2}=\frac{1}{\tau}\int_{-\tau/2}^{+\tau/2}(x_{g}-\varepsilon)^{2}=\frac{\tau^{2}}{2\pi^{2}}
\sum_{k=1}^{\infty}\Phi^{2}(\frac{2\pi k}{\tau})k^{-2}.
\end{equation}

This equation sheds some light on the origin of error $\Delta$ and
defines the conditions on  $\varphi(x)$ which give
$x_{g}=\varepsilon$  . It is evident that $\Delta$  is equal to
zero if $\Phi(2\pi k/\tau)=0$ for $k>0$. Hence, excluding some
very special cases discussed later on, the class of band-limited
functions, with $\Phi(\omega )=0$ for $|\omega |\geq 2\pi/\tau$,
has always $x_{g}$ equal to $\varepsilon$ . Part of this result is
not wholly unexpected. Due to the Wittaker Kotel'nikov Shannon
(WKS) sampling theorem~\cite{9,11}, the subclass of the band-limited
functions, with $\Phi(\omega )=0$ for $|\omega |\geq \pi/\tau$,
has the property that $f_{\epsilon} (x)$ can be exactly
reconstructed from its sampled values $\alpha_{n}(\varepsilon )$
${n=0,\pm1,\pm2,...}$, as can its COG. But the condition for
$\Delta =0$ in Eq. (7) is broader than that of the WKS theorem.
Here, the absence of  overlap of the shifted functions $\Phi(
-2k\pi/\tau)$ $( k =\pm1,\pm2,..)$ with  $\Phi(\omega )$ at
$\omega =0$ suffices, and $x_{g}=\varepsilon$ remains true, even
when the sampling interval must be $\tau/2$ for the WKS theorem.
These properties of $\Delta$ remain valid even if $\varphi(x)$ is
asymmetric. In a broader sense, the error $\Delta$ originates from an
aliasing effect at $\omega =0$ due to the sampling of
$f_{\varepsilon} (x)$ at an overly large interval.

Apart from the mathematical interest of a theorem regarding the
set of sampled functions with  $x_{g}$ equal to $\varepsilon$  ,
the previous condition is of little practical use for our problem.
Other theorems~[9] prove that band-limited functions  $\varphi(x)$
cannot be zero in any finite segment of the $x$-axis, i.e.,
$\varphi (x)$ and $f_{\varepsilon} (x)$ would be different from
zero almost everywhere in x. This property is very far from
standard experimental situations where $\varphi(x)$ is different
from zero only in a narrow region of the $x$-axis covering few
$\tau$'s ($3 \sim 5$ at best). Beyond this region, $\varphi (x)$
rapidly plunges below the readout noise and must be put equal to
zero. Hence, the necessity of keeping the signal well above the
readout noise to increase detection probability  works adversely
to the sensitivity of the subpixel position measurement. The
extreme case of $\varphi (x)$ converging to a Dirac
$\delta$-function is a clear example. This shape maximizes the
detection probability, but leaves its subpixel position completely
undermined, and the mean square error gets its maximum
$\Delta^{2}=\tau^{2}/12$. Simulations will show that practical
situations are intermediate between the full indeterminateness of
the Dirac $\delta$-function and a good reconstruction.

Equation (7) allows the selection of a set of special shapes and
sizes which have  $x_{g}= \varepsilon$ , even if these are not
band-limited. More simply, we have to find functions that have
$\Phi(\omega )=0$ for $ \omega=2k\pi/\tau$ $(k=\pm1,\pm2,\pm3..)$
and $\Phi(0)=1$. It should be pointed out that this distribution
of zeros pertains to the function $\sin(\omega\tau /2)/(\omega)$
and to any of its integer powers and products with functions
regular at $\omega=2k\pi/\tau$. The simplest functions whose FTs
have the above property are the rectangular functions whose sizes
are integer multiples of $\tau$. More complex functions are the
convolutions of identical rectangular functions. These have
special sizes with $x_{g}=\varepsilon$.  For example, triangular
functions (convolutions of two identical rectangular functions)
with sizes that are even multiples of $\tau$ have
$x_{g}=\varepsilon$ . Evidently, this is true for any linear
combination of these special functions and for any convolution
with functions that have FTs regular at $\omega=2k\pi/\tau$. This
class of functions is very large, but the minimum size allowed is
$\tau$ and the convolutions with the minimum-size rectangular
function have sizes $\tau+D$ (where D is the size of the convolved
function).

An application of these considerations  is possible when $\varphi
(x)$ can be expressed as a convolution with a rectangular
function. Now, if the size of the detector coincides  with that of
the rectangular function, the position reconstruction with the COG
algorithm does not require corrections. In all the other cases,
knowledge of $\varphi (x)$ is required to extract $\varepsilon$
given $x_{g}$ as in Eq. (6).

\subsection{ Examples} \indent

From now on, we shall consider only the class of finite support
functions, i.e.,  $\varphi(x)\neq 0$ for $|x|<D/2$. Equation (6)
has the form of an infinite series, but its terms rapidly
decrease, at least as $k^{-2}$, as in the worst case of a
rectangular distribution, and the sum can be cut off after a
reasonable number of terms. In the simulations of Eqs. (6) and
(7), we will explore only four shapes: a rectangle, a triangle, a
cylinder, and a cone. The first two depend solely on $x$; the
second two depend on $x$ and $y$ and must be integrated on $y$ for
our geometry. The rectangle and triangle have easy FTs, while for
the y-integrated cylinder and cone, we obtain the FTs integrating
with the $J_{0}(\omega r)$ Bessel function (Hankel transform).
These simple shapes alone are unable to cover many realistic
conditions. To maximize detection efficiency, a large fraction of
the signal $(70 \sim 80\%)$ is often concentrated on a single
pixel.
\begin{figure}[h!]
\begin{center}
\includegraphics[scale=0.90]{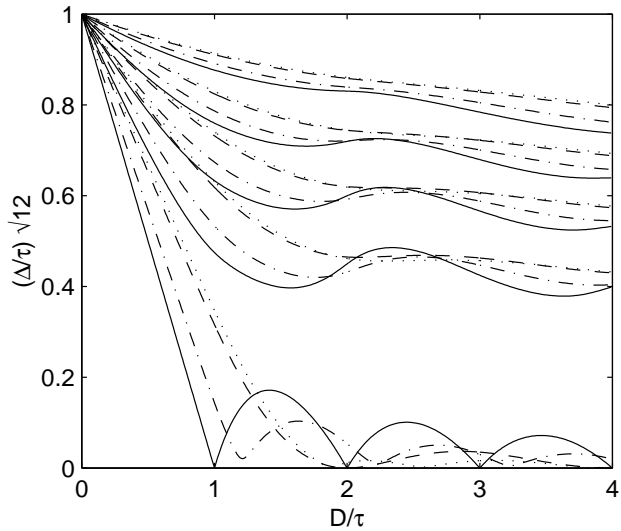}
\end{center}
\caption{\em Plots of $(\Delta/\tau)\surd 12$ versus D/$\tau$ for
the linear combination of two shapes with D'/D=1/20,
$\beta$=0,2,4,8 and four shapes: solid lines for rectangles,
dot-dash lines for cylinders, dashed lines for triangles, and
dotted lines for cones. }
\end{figure}
 To gain some information on the COG error of this kind of
signal distributions, we will simulate them with a linear
combination of two identical functions differing in size D and D':
\begin{equation}
\varphi(x)=[\varphi_{D}(x)+\beta\varphi_{D'}(x)]/(1+\beta)
\end{equation}

The variations in parameter $\beta$ , size D, and ratio D'/D allow
a large increase in the shapes explored. Figure 2) shows  $
(\Delta/\tau)\surd 12$, versus $D/\tau$, for  D'/D=1/20,
$\beta=0,2,4,8$ and the four shapes. Increasing $\beta$ , i.e.,
increasing the signal released in the central pixel, $\Delta$
increases rapidly reaching values almost shape-independent. For
$\beta=0$ (the lowest plots), the sizes for which the rectangular
and triangular shapes have $\Delta =0$ are evident. The
similarities of the cylinder and the cone with the rectangle and
triangle are remarkable.

\begin{figure}[h!]
\begin{center}
\includegraphics[scale=0.9]{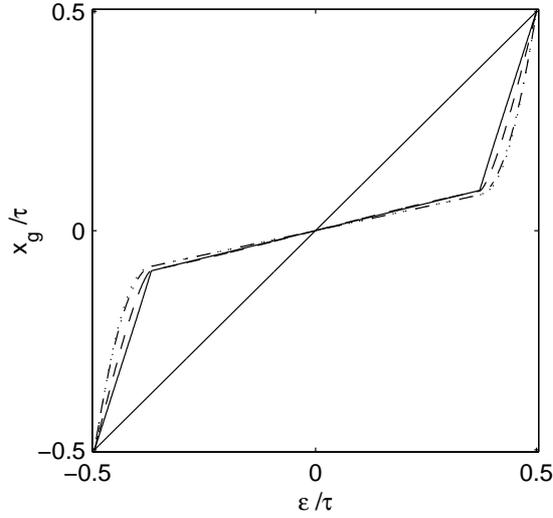}
\end{center}
\caption{\em $x_g$ versus $\varepsilon$ for D=1.8, D'/D=1/7,
$\beta$=3.5, $\tau$=1 and the graphical conventions as in Figure
1. }
\end{figure}

Figure 3)  shows the plot of Eq.(6) for linear combinations whose
$\alpha_{n}(\varepsilon )$ distributions are roughly similar to
those generated on the y-side of the PAMELA silicon tracker~\cite{7} by
a minimum ionizing particle. In principle, this plot  allows the
extraction of $\varepsilon$ given $x_{g}$, or verification of the
optimization of an algorithm.

\section{Histograms}
\subsection{Finite support functions} \indent

To simulate the reconstruction, it is useful to define a general
aspect of the finite support functions  $\varphi(x)$, i.e.,
$\varphi (x)=0$ for $|x|\leq D/2$. The function $\varphi (x)$ can
be expressed as the product of a periodic  function with the
interval function $\Pi(x/D)$ which selects a single period. The
periodic part coincides with $\varphi(x)$ where  $\varphi(x)\neq
0$ and can be expressed as FS. The symmetry simplifies the form
$\varphi(x)$ becoming:
\begin{equation}
\varphi(x)=D^{-1}\Pi(x/D)[\Phi(0)+2\sum_{n=1}^{\infty}\Phi(\gamma_{n})\cos(\gamma_{n}x)].
\end{equation}

The parameters $\Phi(\gamma_{n})$ are the values of
$\Phi(\omega)$, the FT of $\varphi(x)$, calculated at the
FS-frequencies $\gamma_{n}=2\pi n/D$, and the dual form of the WKS
theorem express the form of $\Phi(\omega)$ with the parameters
$\Phi(\gamma_{n})$. The normalization  fixes  $\Phi(0)=1$ for any
D, so $\varphi(x)$ must scale on D as
$\varphi(x)\rightarrow\varphi(x/D)D^{-1}$ and the parameters (form
factors) $\Phi(\gamma_{n})$ turn out to be $D$-independent. This
property is very convenient for our use of a linear combination of
identical functions with different support $D$ as in Eq. (8).
Here, only a set of $\Phi(\gamma_{n})$ must be calculated, and the
cutoff on the high frequencies of  $\Phi(\omega)$ is identical for
the two (or more) functions which form $\varphi(x)$.

For signal distributions scaling with a length $L_{M}$ typical of
the material, Eq.(9) has another interesting property: $D$ scales
with $L_{M}$, and the set {$\Phi(\gamma_{n})$} is independent of
the material, and can be defined  once and for all. An example of
this  is the average photon shower and its lateral scaling with
the Molier radius~\cite{5}.

\subsection{Histograms} \indent

Using Eqs.(1) and (9), the function $f(x)$ can be calculated
explicitly, and  for $D>\tau$ becomes:
\begin{align}
f(x)=&\frac{1}{D}\Big\{\tau+g_{+}(x)[x+\frac
{D-\tau}{2}]+g_{-}(x)[x-\frac {D+\tau}{2}]\Big\}+ \nonumber\\
&\sum_{n=1}^{\infty}\frac{1}{\pi
n}\Big\{\sin[\gamma_{n}(x+\frac{\tau}{2})][1-g_{-}(x)]-\sin[\gamma_{n}(x-\frac{\tau}{2})][1-g_{+}(x)]\Big\}\Phi(\gamma_{n})
\end{align}
where  $g_{\pm}(x)=\Pi((2x\pm D)/2\tau)$ and $x$ is limited by
$-(D+\tau)/2< x <(D+\tau)/2$. Outside this interval, $f(x)$ is
zero. For $D\leq\tau$, $f(x)$ equals Eq. (10), excluding the
region $(-\tau+D)/2 < x < (\tau-D)/2$ where $f(x)=1$. In Eq. (10),
functions $g_{\pm}(x)$ are introduced to write in compact form the
changes due to different integration limits. The tradeoff for
compactness is difficulty for the differentiating.

The sampling of $f(x)$ at points $x_{m} =m\tau-\varepsilon$  (for
$m =\pm 1,\pm2,..$) generates the set
$\{\alpha_{m}(\varepsilon)\}$. With a realistic noise model, Eq.
(10) allows the simulation of the experimental data given the form
factors $\Phi(\gamma_{n})$, or it allows the extraction of the
form factors $\Phi(\gamma_{n})$, given a value of $D$, from a
distribution of experimental data (as done in a slightly different
way in ~\cite{12}). For construction, the samples of $f(x)$ at points
$x_{m}=m\tau-\varepsilon$ generate the histogram of $\varphi(x)$
with bin-size $\tau$. The generalization of Eq.(10) to an
asymmetric function $\varphi(x)$ is simple. Quite often, the fit
to histogram is done directly with $\varphi(x)$; assuming its
value proportional to the value at the center of each histogram's
bin, it is evident that $f(x)$ (not $\tau\varphi(x)$) has this
property. This inconsistency almost always disappears as the
bin-size goes to zero for regular functions.

Improved centroid algorithms can be obtained  by fitting $f(x)$
with a few parameter functions and extracting $f(x-\varepsilon )$
and $\varepsilon$  from the experimental data as done in~\cite{10} in a
special case. These algorithms are often called "nonlinear COGs"~\cite{13},
but their extraction from a set of predefined solutions as
experimented in~\cite{13} looks very improbable. Matching case by case
is probably the only way to proceed.

\subsection{Sum of the series} \indent

In Eq. (10),  the infinite series have terms decreasing as fast as
$n^{-(|m|+1)}$ if $m$ is the degree of continuity of $\varphi(x)$,
Hence, not too many terms are required to arrive at a significant
result. Equations (6) and (7) require $\Phi(2n\pi/\tau)$ values
differing from the form factors $\Phi(\gamma_{n})$. The connection
between these two sets is given by the dual  formulation of   the
WKS theorem which states that the FT of Eq. (9), $\Phi(\omega)$,
is given by:
\begin{equation}
\Phi(\omega)=\sum_{n=-\infty}^{+\infty}\Phi(\gamma_{n})\frac{\sin(\omega
D/2-n\pi)}{(\omega D/2-n\pi)}.
\end{equation}
This procedure can be quite cumbersome in cases such as axially
symmetric bidimensional shapes which often have complicated FTs.
In addition, for signal distributions that scale with a length
typical of the material, the set of values $\{\Phi(\gamma_{n})\}$
is material independent, and hence it is important to express the
COG systematic error by those values. For this task, we found a
way of summing the $k$-dependence of  Eq. (6) for any
$\varepsilon$ and $D$. With the general form of $\Phi(\omega)$
given by the WKS theorem, the $k$-series for $x_{g}$  becomes:
\begin{align}
x_{g}=\varepsilon-\frac{\tau}{4\pi}{\sum_{k=-\infty}^{+\infty}}_{k\neq
0}\frac{(-1)^{k}}{k}
\sum_{n=-\infty}^{+\infty}\Phi(\gamma_{n})\frac{(-1)^{n}\big[e^{i2k
\pi(\varepsilon+D/2)/\tau}-e^{i2k\pi(\varepsilon-D/2)/\tau}\big]}{(kD\pi/\tau-n\pi)}.
\end{align}
Exchanging the order of the sums, we can recast the sum on $k$ as
an integration along a closed line~\cite{14} multiplied by an
integration function $[\sin(\pi z)]^{-1}$ that has poles for
integer $z$-values:
\begin{align}
x_{g}=\varepsilon-\frac{\tau}{4\pi}\sum_{n=-\infty}^{+\infty}(-1)^{n}\Phi(\gamma_{n})
\oint_{C_{1}}dz\frac{1}{z\sin(\pi z)}\frac{\big[e^{i2k
\pi(\varepsilon+D/2)/\tau}-e^{i2k\pi(\varepsilon-D/2)/\tau}\big]}{(\frac{\pi
zD}{\tau}-n\pi)},\tag{12'}\nonumber
\end{align}
$C_{1}$ is a closed path which encircles all the poles of
$[\sin(\pi z)]^{-1}$, but excludes the $z=0$ pole and all the
poles of $\Phi(\omega)$. The coincidence of the $\Phi(\omega)$
poles with that of $[\sin(\pi z)]^{-1}$ can be avoided by adding a
small imaginary part to the denominators of $\Phi(\omega)$ and
keeping the limit at the end of the sum. Path $C_{1}$ can be
deformed toward an infinite circle, and, if the integral goes to
zero on this path, the sum of the residues at $z=0$ and at the
poles of $\Phi(\omega)$ gives the sum of the series. The integral
is zero on the infinite circle if $|(2\varepsilon\pm
D)/2\tau|\leq1/2$ i.e., in this case the exponents in the
numerator of Eq.(12) are smaller or equal to that of the
$[\sin(\pi z)]^{-1}$. This limitation on $\varepsilon$ and $D$,
however, is too restrictive to render the summation useful. We
need the sum for values of $\varepsilon$ and $D$ as large as they
may be. To overcome this, we can observe that, in many cases, the
WKS theorem could not be used for $\Phi(\omega)$, since a large
set of candidates  $\varphi(x)$ has a simple closed FT. For these
functions, at the $\varepsilon$ value for which $|(2\varepsilon\pm
D)/2\tau|=1/2$, the $k$-sum is regularly convergent. These limits
simply indicate the point where a second strip starts to collect
signals. Hence, with the sum being convergent, we can try to
remove the limitations to the residue theorem. We can observe that
the contribution to $x_{g}$ in Eq. (12) from the two exponents
$\xi_{\pm}=(2\varepsilon\pm D)/2\tau$ is through periodic function
with period one, and adding or subtracting an integer number to
$\xi_{\pm}$ in the exponents leaves the sum of the series without
variation. This allows the substitution of $\xi_{\pm}$ with
$\xi_{\pm}+m_{\pm}$ in Eq. (12) where $m_{\pm}$ are integers that
always give $|\xi_{\pm}+m_{\pm}|\leq1/2$. Now the integral in Eq.
(12') can be calculated with the residue theorem for any value of
$\varepsilon$ and $D$. To give a functional form to the previous
transformation of $\xi_{\pm}$, it is convenient to use the
function
$\Theta(\xi_{\pm})=\xi_{\pm}-\mathrm{floor}(\xi_{\pm}+1/2)$. This
is a sawtooth function, with $|\Theta(\xi_{\pm})|\leq 1/2$. Note
that any other definition of a sawtooth function works
identically. Our form is very convenient in computer usage, but it
is not suited for derivation. The sum of the series in Eq. (12)
becomes:
\begin{align}
&x_{g}=\varepsilon+\frac{\tau^{2}}{2D}\big[\Theta^{2}(\xi_{-})-\Theta^{2}(\xi_{+})\big]+
\frac{\tau}{2\pi}\sum_{n=1}^{+\infty}\frac{(-1)^{n}\Phi(\gamma_{n})}{n\sin(\gamma_{n}\tau/2)}
\big\{\cos[\gamma_{n}\tau\Theta(\xi_{+})]-\cos[\gamma_{n}\tau\Theta(\xi_{-})]\big\}\\
&\xi_{\pm}=(2\varepsilon\pm D)/2\tau\nonumber\\
&\Theta(\xi_{\pm})=\xi_{\pm}-\mathrm{floor}(\xi_{\pm}-1/2)\nonumber
\end{align}

Here $x_{g}$ is linear in the form factors $\Phi(\gamma_{n})$ and
is suited for a best fit to extract  $\varphi(x)$ from the
measured function $x_{g}(\varepsilon)$. Equation (13) gives an
explicit and simple form for a rectangular shape where all the
$\Phi(\gamma_{n})$'s are zero. For an assembly of two rectangular
functions, $x_{g}$ is:
\begin{equation}
x_{g}=\varepsilon+\frac{\tau^{2}}{2(1+\beta)}\Big\{\big[\Theta^{2}(\frac{2\varepsilon-D}{2\tau})-
\Theta^{2}(\frac{2\varepsilon+D}{2\tau})\big]\frac{1}{D}+\big[\Theta^{2}(\frac{2\varepsilon-D'}{2\tau})-
\Theta^{2}(\frac{2\varepsilon+D'}{2\tau})\big]\frac{\beta}{D'}\Big\}.\nonumber
\end{equation}

An assembly of identical functions, scaled with D as in Eq.(8),
has the same $\Phi(\gamma_{n})$ terms, and they factor out. We can
see here a reason for $\varphi(x)$ in the form of  Eq. (8), and
how important the proper selection of the ranges D and D' (or
more) is. The number of $\Phi(\gamma_{n})$ to be used in Eq. (13)
can be drastically reduced.

\section{A more complex setup}
\subsection{Generalization of the spatial integrator}\indent

With few modifications, these approaches can be extended to a
detector set that does not operate on the signal as a perfect
spatial integrator. This is a frequent case: For example, the
crystal detectors of an em calorimeter do not collect the light
with constant efficiency over all their sizes. They have signal
losses at their borders due to mechanical tolerances or to
nonhomogeneous light  absorption at the borders, with the
effective efficiency modulated over the crystal.

In silicon strip detectors, the signal collection is even more
complex. Some setups have unread strips which spread the signal by
capacitive couplings among nearby strips as in the AMS~\cite{14} or
PAMELA~\cite{7} trackers. To handle these situations, Eq. (1) must be
generalized:
\begin{align}
&\mathrm{f}(x)=\int_{-\infty}^{+\infty}\mathrm{p}(x-x',\tau_{1})\varphi(x')dx'\\
&\mathrm{f}_{\varepsilon}(x)=\int_{-\infty}^{+\infty}
\mathrm{p}(x-x',\tau_{1})\varphi(x'-\varepsilon)dx' \nonumber
\end{align}
Now $\mathrm{p}(x,\tau_{1})$ is a generic response function
symmetric around zero with a finite range $\tau_{1}$ (the sampling
distance is always $\tau$). For $\tau_{1}<\tau$, there is a
definite signal loss; for $\tau_{1}>\tau$ there is a long range
coupling (crosstalk) and the signal collected by a strip modifies
the signal collected by the nearby strips. From Eq.(14),
$\mathrm{p}(x,\tau_{1})$ is the response function of the detector
to a Dirac $\delta$-signal. Due to its finite range and symmetry,
the form of $\mathrm{p}(x,\tau_{1})$ resembles Eq. (9):
\begin{equation}
\mathrm{p}(x,\tau_{1})=\Pi(x/\tau_{1})[G_{0}+2\sum_{n=1}^{+\infty}G_{n}\cos(2\pi
x/\tau_{1})].
\end{equation}
Its FT is given by the WKS theorem:
\begin{equation}
\mathrm{P}(\omega,\tau_{1})=\tau_{1}\sum_{n=-\infty}^{+\infty}G_{n}\frac{\sin(\omega\tau_{1}/2-n\pi)}
{(\omega\tau_{1}/2-n\pi)}.
\end{equation}
Now $\mathrm{F}_{\varepsilon}(\omega)$ becomes:
\[
\mathrm{F}_{\varepsilon}(\omega)=\tau_{1}\sum_{n=-\infty}^{+\infty}G_{n}\frac{\sin(\omega\tau_{1}/2-n\pi)}
{(\omega\tau_{1}/2-n\pi)}\Phi(\omega)\mathrm{e}^{-i\varepsilon\omega}.
\]
A simple case is that with $G_{0}=1$, $G_{n}=0$ for $n\neq 0$ and
$\tau_{1}<\tau$. Here, each detector  is smaller than the sampling
interval, i.e., there is a complete loss of  signal at the borders
of two nearby detectors due to a hole with size $\tau-\tau_{1}$.
The efficiency $\mathrm{S}_{\varepsilon}(0)$ of the set of
detectors is less than one for this loss:
\[
\mathrm{S}_{\varepsilon}(0)=\frac{\tau_{1}}{\tau}+\sum_{k=1}^{+\infty}
\frac{\sin(k\pi\tau_{1}/\tau)}{k\pi}\Phi(-2k\pi/\tau)\cos(2\pi\varepsilon
k/\tau). \]\\
Now, the sum of all the signals collected by the array of
detectors acquires an explicit dependence from the impact point.
This $\varepsilon$ dependence is periodic with period $\tau$ and
symmetric respect to $\varepsilon=0$. Efficiencies of this form
are well known in em-calorimeters.
\subsection{Crosstalk}\indent

Crosstalk is a very common effect which must be carefully
simulated. We will derive special crosstalk shapes whose $x_{g}$
are free of discretized error for any signal distribution. Unlike
signal distribution where few controls are left to the user,
crosstalk may be controlled by the detector fabrication and hence
it may be optimized. For example, the unread strips in a silicon
strip detector are steps in this direction. To discuss in depth
these further effects, we need the forms of  $x_{g}$ and
$\mathrm{S}_{\varepsilon}(0)$ with the explicit dependence from
the generalized response function $\mathrm{p}(x,\tau_{1})$, that
describes crosstalk for $\tau_{1}>\tau$.  We can write $x_{g}$ as:
\begin{equation}
x_{g}=\varepsilon-\frac{2}{\tau\mathrm{S}_{\varepsilon}(0)}
\sum_{k=1}^{+\infty}[\Phi(-2k\pi/\tau)\mathrm{P}'(-2k\pi/\tau,\tau_{1})+\Phi'(-2k\pi/\tau)\mathrm{P}(-2k\pi/\tau,\tau_{1})]
\sin(2k\pi\varepsilon/\tau)
\end{equation}
where $\Phi'(-2k\pi/\tau)$ and $\mathrm{P}'(-2k\pi/\tau,\tau_{1})$
are the derivatives of $\Phi(\omega)$ and
$\mathrm{P}(\omega,\tau_{1})$ with respect to $\omega$ calculated
for $\omega=-2k\pi/\tau$. $\mathrm{S}_{\varepsilon}(0)$ is given
by:
\[
\mathrm{S}_{\varepsilon}(0)=\frac{1}{\tau}\sum_{k=-\infty}^{+\infty}\mathrm{P}(-2k\pi/\tau,\tau_{1})
\Phi(-2k\pi/\tau)\cos(2k\pi\varepsilon/\tau).\tag{17'}
\]
Let us consider first when $\mathrm{S}_{\varepsilon}(0)=1$ for any
$\varepsilon$. We will call this condition lossless crosstalk, or
uniform crosstalk. It is clear that the condition
$\mathrm{S}_{\varepsilon}(0)=1$ models a crosstalk which spreads
the signal among various detectors, but saves the total signal.
The simplest setup with $\mathrm{S}_{\varepsilon}(0)=1$ was
encountered in Section 2.3) with
$\mathrm{p}(x,\tau_{1})=\Pi(x/\tau)$. Although it has no
crosstalk, it serves to illustrate the meaning of uniform
crosstalk. If we examine the neighboring regions of efficiency of
the  $\Pi(x/\tau)$ detector, the total efficiency is a constant
function (almost everywhere)
$\sum_{n=-\infty}^{+\infty}\Pi(x-n\tau)=1$. We will prove in
Sections 5.5 and 5.6 that functions $\mathrm{p}(x,\tau_{1})$ such
that $\sum_{n=-\infty}^{+\infty}\mathrm{p}(x-n\tau)=1$
(generalization of the property of the interval function) have
uniform crosstalk. For now, we can observe that the condition
$\mathrm{S}_{\varepsilon}(0)=1$ in Eq.(17') is assured by
$\mathrm{P}(-2k\pi/\tau,\tau_{1})=0$ for $k=\{\pm 1,\pm 2,\pm
3,...\}$ and $\mathrm{P}(0,\tau_{1})=\tau$. These conditions are
identical to those in of Section 3.1 for the absence of the
discretization error in Eq. (6). The only difference is the
normalization of $\mathrm{p}(x,\tau_{1})$ which is $\tau$, when
$\varphi(x)$ is normalized to one. Therefore, the
$\mathrm{p}(x,\tau_{1})$ functions with uniform crosstalk are all
those which give $x_{g}=\varepsilon$ in Eq. (6) when used in Eq.
(1) in the place of $\varphi(x)$. A consequence of this, for
$\tau<\tau_{1}<2\tau$ the $\mathrm{p}(x,\tau_{1})$ functions with
uniform crosstalk are convolutions of normalized functions of
range D with $\Pi(x/\tau)$ $(D+\tau=\tau_{1})$. Starting from
$\tau_{1}\geq2\tau$, other special
$\mathrm{p}(x,\tau_{1})$-functions will be isolated, in addition
to the one discussed above. From Eq. (16), it is clear that, for
$\tau_{1}=2\tau$ and $G_{n}=0$ for $n=2k$ and $G_{0}\neq0$, a set
of uniform crosstalk is generated. This class of functions
contains all the uniform crosstalk functions with the range
$\tau_{1}=2\tau$. The functions described above as convolutions
with interval functions are in this class if $D+\tau=2\tau$. The
convolution of functions of this class with range $D$ functions
are uniform crosstalk functions with range $D+2\tau$, and so on
with $\tau=3, 4,..$. Crosstalk models for silicon strip detectors
can be extracted from this class of uniform crosstalk functions.

In the presence of uniform crosstalk, Eq. (17) simplifies, the
last term in square brackets disappears and Eq. (17) becomes
similar to Eq. (6):
\[
x_{g}=\varepsilon-\frac{2}{\tau}
\sum_{k=1}^{+\infty}\Phi(-2k\pi/\tau)\mathrm{P}'(-2k\pi/\tau,\tau_{1})
\sin(2k\pi\varepsilon/\tau)\tag{17''}
\]
This equation suggests another strategy for eliminating the
discretization error. If the first derivative of the FT of the
crosstalk function pertains to the class of the uniform crosstalk
functions, the COG of the signals collected by the set of
detectors coincides with the COG of the signal distribution for
any signal distribution. The easiest crosstalk function with this
property is the triangular function with range $D=2\tau$. The FT
of a triangular function is the square of the FT of an interval
function  of range $D/2$ and has double zeros at $\omega=4k\pi/D$
 $(k=\pm 1,\pm 2,\pm 3 ..)$ and its derivative has simple zeros in the same
points, so for $D=2\tau$ the sum in Eq. (17'') is zero. It is
evident that  crosstalk functions, convolutions of finite range
functions with a triangular function with range $2\tau$, eliminate
the discretization error in Eq.(17'').  The properties of the
crosstalk functions are more interesting than those regarding the
shapes of the signal distributions. The signal distributions are
rarely modifiable, unlike the crosstalk functions that could be
tuned to achieve a triangular shape. For example, a probable step
in this direction could be the silicon strip detector used in  AMS
[15], but detailed simulations must be performed to test how near
these detectors are to ideal triangular crosstalk.

Getting back to a generic crosstalk, we can even write the
explicit form of the  function f$(x)$ in this  case. The
integration ranges are identical to those of Eq. (10), but now the
products of the form $\cos(2\pi mx/\tau_{1})*\cos(2\pi nx/D)$ must
be integrated. Since the resulting equation is too long to be
given here,  we will give in the following an alternative form
that is more transparent, compact, and easy to use.

The modulation of the signal, due to  p$(x,\tau_{1})$, adds new
unknowns to the problem. These must be extracted by the physical
properties of the detector and reformulated as a response to a
Dirac $\delta$-signal to be inserted in Eq. (14). If this
modulation can be reduced to an assembly of rectangular functions,
Eq.(10) can be used directly for each one. The resulting f$(x)$
will be a linear combination of a set of Eq.(10). Each member is
calculated with a proper $\tau_{i}$ and multiplied by the
amplitude of the corresponding rectangular function. The sampling
of f$(x-\varepsilon)$ at interval $\tau$ gives the set
$\{\alpha_{n}(\varepsilon)\}$. It is possible to show that this
arrangement, with a few rectangular functions, allows the
simulation of the signal spread by a capacitive coupling in micro
strip silicon detectors with unread strips.

\subsection{Finite set of sensors}\indent

Up to now, we have considered an infinite array of identical
detectors, each one accounting for the signal released upon it.
This produced the $\varepsilon$-periodicity for $x_{g}$ and
S$_{\varepsilon}(0)$. The cutoff was given by the finite ranges
$D$ and $\tau_{1}$ of $\varphi(x)$ and p$(x,\tau_{1})$.

In data analysis, the low-signal detectors are often suppressed
for energy and position reconstruction, which increases immunity
to noise at the expense of a loss in information. For example, in
{em}-calorimeters, the energy is reconstructed from the signal
detected in a fixed number of crystals around the center of the
cluster; 3x3 or 5x5 clusters are standard choices in the CMS
em-calorimeter~\cite{4}.  The cuts, while  reducing noise and
fluctuations, introduce an additional dependence on the position
of the photon impact point  in  the reconstructed energy. This can
be understood by observing that, as the impact point nears the
edge of the central sensor, very different tails are suppressed to
the right  as compared to those suppressed to the left. Now, there
is generally no compensation between what is lost on one hand and
what is gained on the other.  Equation (10) evidently simulates
this cut, but no explicit analytical dependence on $\varepsilon$
can be extracted.  To obtain  analytical expressions of $x_{g}$
and of S$_{\varepsilon}(0)$, we must  modify to some extent
Eqs.~(2) and (2'). There, the sampling  extends  from $-\infty$ to
$+\infty$, completely covering any range of the function
f$(x-\varepsilon)$, thereby enabling us to automatically treat
band-limited functions that are not finite range. This form of
sampling, however, is too broad for our limitation to finite
ranges for  f$(x-\varepsilon)$ which is, at the most,
$D+\tau+\tau_{1}$ a range that is often less than a few times
$\tau$ $(3\sim5\tau)$. A key property of  Eq.(2) is the
possibility to explicitly calculate the convolution in
$S_{\varepsilon}(\omega)$ through the Poisson identity.
Limitations on the sampling number introduce a multiplication by
auxiliary interval functions to suppress the tails of
$f(x-\varepsilon)$. Thus, Eqs. (2) and (2') become:
\begin{align}
&\mathrm{s}_{\varepsilon}(x)=\sum_{n=-M}^{+M}\alpha_{n}(\varepsilon)\delta(x-n\tau)\nonumber\\
&\mathrm{s}_{\varepsilon}(x)=\mathrm{f}(x-\varepsilon)\Pi(x/\Lambda)
\sum_{n=-\infty}^{+\infty}\delta(x-n\tau)\nonumber
\end{align}
Now, the Poisson identity can be used on the convolution of
F$_{\varepsilon}(\omega)$ with the FT of $\Pi(x/\Lambda)$, where
$\Lambda=(2M+1)\tau$ is the region of true sampling. The FT of the
interval function is easy, but its convolution with
F$_{\varepsilon}(\omega)$ has no closed analytical expression. We
can get around this by applying the following observations: Due to
the limits on the sampling ranges, from $-2\tau$ to $+2\tau$ at
the most, we can neglect the specific form of f$(x-\varepsilon)$
far from our sampling region. We can use a periodic form,
indicated by $\varphi^{p}(x)$ and f$^{p}(x)$, of $\varphi(x)$ and
of f$(x)$ with an arbitrary period T larger than the sampling
region, and coinciding with $\varphi(x)$ and f$(x)$ on a period.
The sampling will test only a part of the period $-T/2<x<T/2$ of
f$^{p}(x)$ and its result will be identical to the sampling of
f$(x)$. Due to the periodicity of $\varphi^{p}(x)$ and f$^{p}(x)$,
their FT is a sum of Dirac $\delta$-function, and  the additional
convolution can be analytically treated. The periodic function
$\varphi^{p}(x)$ is defined through the FS:
\begin{equation}
\varphi^{p}(x)=\sum_{n=-\infty}^{+\infty}\Phi_{n}^{p}\ \exp(i2\pi
nx/T),
\end{equation}
with the Fourier components $\Phi_{n}^{p}$ given by $(T>D)$:
\begin{equation*}
\Phi_{n}^{p}=\frac{1}{T}\int_{-D/2}^{+D/2}\mathrm{d}x\varphi(x)\exp(-i2\pi
nx/T)
\end{equation*}
For its construction, $\varphi^{p}(x)$ coincides with $\varphi(x)$
in a period, and $T\Phi_{n}^{p}=\Phi(2n\pi/T)$. For the Poisson
identity~\cite{9}, it is, as expected:
\begin{equation}
\varphi^{p}(x)=\sum_{n=-\infty}^{+\infty}\Phi_{n}^{p}\ \exp(i2\pi
nx/T)=\sum_{k=-\infty}^{+\infty}\varphi(x-kT)
\end{equation}
Now, $\Phi^{p}(\omega)$, the FT of $\varphi^{p}(x)$, is:
\begin{equation}
\Phi^{p}(\omega)=\frac{2\pi}{T}\sum_{n=-\infty}^{+\infty}\Phi(\frac{2n\pi}{T})\delta(\omega-\frac{2n\pi}{T}).
\end{equation}
Due to the finite support of $\varphi(x)$, the function
$\varphi^{p}(x)$ is zero for $-T/2\leq x<-D/2$ and $D/2>x\geq
T/2$. This indicates some practical complications of the approach.
The zero regions are generated by the interference of the
high-frequency components of $\varphi^{p}(x)$. To achieve this, we
need more terms in the FS than required in Eq.(9). Generally,
since a large $T$ implies a relatively larger number of
frequencies to be handled, it is better to use the smallest
possible $T$ value: In practice $\varphi^{p}(x)$ is used in a
convolution with p$(x,\tau_{1})$, and this is a low-pass filter
which heavily suppresses the high-frequency components.
Nevertheless, the computer usage of these and the following
equations is as easy as the previous ones, or easier. The
simulations are contained in a few lines of MATLAB code.

The function  f$^{p}(x)$ is the convolution of the periodic
function $\varphi^{p}(x)$ with the finite-range function
p$(x,\tau_{1})$, It is periodic as $\varphi^{p}(x)$ and coincides
with f$(x)$ on a period if $T>D+\tau_{1}$. For periods $T$, below
$D+\tau_{1}$, the tails of f$^{p}(x)$ differ from f$(x)$. But, due
to the reduced number of samples, these differences do not matter
provided they are outside the sampling regions. With Eq. (20),
f$^{p}(x)$ becomes:
\begin{equation}
\mathrm{f}^{p}(x)=\sum_{n=-\infty}^{+\infty}\Phi_{n}^{p}\
\mathrm{P}(\frac{2n\pi}{T},\tau_{1})\exp(\frac{i2\pi nx}{T}).
\end{equation}
Equation (21) is the promised generalization of Eq. (10) for any
p$(x,\tau_{1})$.

The sampling of f$^{p}(x-\varepsilon)$ in three points (for
$-\tau/2<\varepsilon<\tau/2$) is given by:
\[
\mathrm{s}_{\varepsilon}(x)=\mathrm{f}^{p}(x-\varepsilon)[\delta(x+\tau)+\delta(x)+\delta(x-\tau)].
\]
The function S$_{\varepsilon}(\omega)$, the FT of
s$_{\varepsilon}(x)$, is the convolution of
F$^{p}(\omega)e^{-i\omega\varepsilon}$, FT of
f$^{p}(x-\varepsilon)$, with the FT of the sampling function
H$(\omega)$. Now, S$_{\varepsilon}(\omega)$ has the form:
\begin{equation}
\mathrm{S}_{\varepsilon}(\omega)=\sum_{n=-\infty}^{+\infty}\Phi_{n}^{p}\
\mathrm{P}(\frac{2n\pi}{T},
\tau_{1})\exp(\frac{-i2n\pi\varepsilon}{T})\mathrm{H}(\omega-\frac{2n\pi}{T}).
\end{equation}
The function H$(\omega)$ is generally expressed by:
\[
\mathrm{H}(\omega)=\sum_{n=-M_{1}}^{+M_{2}}\exp(-i\omega n\tau).
\]
where $M_{1}$ and $M_{2}$ are the leftmost and the rightmost
sampling points.

Turning back to Eq. (22), we have to recall that our decision to
render periodic $\varphi(x)$ and f$(x)$, is a question of taste.
Likewise, we can make the sampling function  periodic, and keep
$\varphi(x)$ and f$(x)$ untouched. Even the latter is finite
ranged, and its form does not matter outside the  f$(x)$ range. In
these assumptions, the equation for $x_{g}$ implies derivation of
F$_{\varepsilon}(\omega)$, and these look more complex than  Eq.
(22) and its coming derivations.

Equations (21) and (22) could be handled according to the
procedures in Section 4.3, but the resulting expressions would be
overly long and complex.

\subsection{Two, three and more-than-three sensors}

For three and five sampling points H$(\omega)$ is:
\begin{align*}
&\mathrm{H}(\omega)=1+2\cos(\omega\tau)\qquad\qquad\qquad\ \ \mathrm{3 points}\nonumber\\
&\mathrm{H}(\omega)=1+2\cos(\omega\tau)+2\cos(2\omega\tau)\quad\mathrm{5
points}\nonumber
\end{align*}
With three samples, such as in the case of three crystal rows, the
signal collected is:
\begin{equation}
\mathrm{S}_{\varepsilon}(0)=\frac{3\
\mathrm{P}(0,\tau_{1})}{T}+2\sum_{n=1}^{+\infty}\Phi_{n}^{p}\
\mathrm{P}(\frac{2n\pi}{T},\tau_{1})\cos(\frac{2n\pi\varepsilon}{T})\big[
1+2\cos(\frac{2n\pi\tau}{T})\big],
\end{equation}
and the center of gravity $x_{g}$ becomes:
\[
x_{g}=\frac{4}{\mathrm{S}_{\varepsilon}(0)}\sum_{n=1}^{+\infty}\Phi_{n}^{p}\
\mathrm{P}(\frac{2n\pi}{T},\tau_{1})\sin(\frac{2n\pi\tau}{T})\sin(\frac{2n\pi\varepsilon}{T}).\tag{23'}
\]
It is evident that S$_{\varepsilon}(0)$ has an explicit
$\varepsilon$-dependence even with the function
P$(\omega,\tau_{1})$ of Equations (5) and (15), which gives
S$_{\varepsilon}(0)=1$ for infinite sampling. In the following, we
will discuss a typical defect of the three sampling points COG.

With slight modification of H$(\omega)$, we are able to calculate
the properties and the details of the two strips' COG. This method
is widely used for position reconstruction in silicon micro strip
detectors [7,15,16].  The limitation to two strips  is  viable for
noise reduction, but, if improperly used, entails a large
systematic error as shown in Figure 4. The standard application
avoids the effects with the introduction of the so-called $\eta$
response function~\cite{16}. The  $\eta$-function is, by definition,
the ratio $Q_{L}/(Q_{L}+Q_{R})$, where $Q_{L}$ and $Q_{R}$ are the
left and right strips of the two largest signal couples. It is
evident that the $\eta$-function is directly connected to the two
strips' COG, and here $x_{g}$ has a discontinuity at
$\varepsilon=0$  for $D+\tau_{1}>2\tau$, as is often the case. The
two strips' COG is very asymmetric, but around $\varepsilon=0$ the
signal symmetrically distributes in the strips to the left and to
the right of the central strip. Hence, the suppression of the
signal in one of the two moves the value of  $x_{g}$ by a fixed
amount in the other direction, thereby generating the
discontinuity. The calculation of $x_{g}$ and its discontinuity is
straightforward from Equations (22) and (3). Two different
functions H$(\omega)$ must be considered:
\begin{align*}
&\mathrm{H}(\omega)=1+\mathrm{e}^{(-i\omega\tau)}\qquad
\varepsilon>0\\
&\mathrm{H}(\omega)=1+\mathrm{e}^{(+i\omega\tau)}\qquad\varepsilon<0
\end{align*}
The total signal collected by two strips $(Q_{L}+Q_{R})$ is
continuous in $\varepsilon=0$:
\[
\mathrm{S}_{\varepsilon}(0)=\frac{2\mathrm{P}(0,\tau_{1})}{T}+2\sum_{n=1}^{+\infty}\Phi_{n}^{p}\
\mathrm{P}(\frac{2n\pi}{T},\tau_{1})\big[\cos(\frac{2n\pi\varepsilon}{T})+\cos(\frac{2n\pi(|\varepsilon|-\tau)}{T})\big],
\]
and $x_{g}(\varepsilon)$ becomes:
\[
x_{g}=\frac{\mathrm{sign}(\varepsilon)\tau}{\mathrm{S}_{\varepsilon}(0)}\big[\frac{\mathrm{P}(0,\tau_{1})}{T}+2\sum_{n=1}^{+\infty}
\Phi_{n}^{p}\
\mathrm{P}(\frac{2n\pi}{T},\tau_{1})\cos(\frac{2n\pi(|\varepsilon|-\tau)}{T})\big].
\]
The discontinuity is quite evident, and can be extracted from the
above equations: Expressed by  f$(x)$, $\Delta x_{g}=2\tau
\mathrm{f}(\tau)/(\mathrm{f}(0)+\mathrm{f}(\tau))$. For
$D+\tau_{1}<2\tau$, f$(\tau)$ is zero, and the discontinuity
disappears. From $x_{g}$, we can recalculate $\eta$, and the
probability of having a $\eta$-value. In the histograms of the
$\eta$-distribution, the discontinuity is signaled by the presence
of zero probability (or better a probability drop) for values of
immediately above zero and below one. In the histogram of the
$x_{g}$ distribution, the zero probability region is located, as
can be expected, around $x_{g}=0$ (Figure~4).

%
\begin{figure}[h!]
\begin{center}
\includegraphics[scale=0.8]{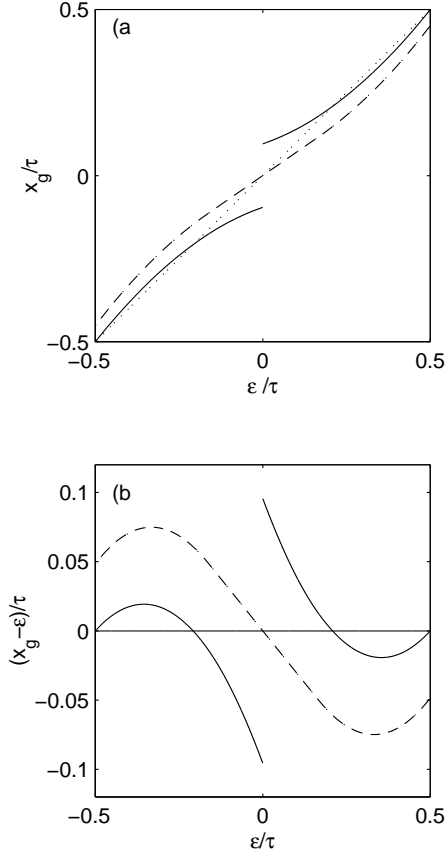}
\end{center}
\caption{\em a), Plots of $x_g$ versus $\varepsilon$ for a linear
combination of two triangles with sizes D=4, D'/D=1/3, $\beta=3$,
solid line for a two-sensor COG, dot-dash line for a three-sensor
COG, and dotted line for $x_g=\varepsilon$. The  discontinuities
at $\varepsilon=0$ for the two-sensors COG, and at
$\varepsilon=\pm0.5\tau$ for three-sensor COG are clearly visible.
In b), the plots of the $(x_g-\varepsilon)/\tau$ versus
$\varepsilon/\tau$ are shown for both COG, here the
discontinuities are more visible. }
\end{figure}
%
The three-sensor COG has discontinuities if the range of f$(x)$
(which is $D+\tau_{1}$) is larger than  $3\tau$. The
discontinuities are located at $\varepsilon=\pm\tau/2$ and often
go unnoticed. The points $\varepsilon=\pm\tau/2$ signal the
transition from one set of three sensors to another. When
$\varepsilon$ approaches $\tau/2$, i.e., the border of the central
detector, the signal distribution tends to be symmetric with
respect to this point, but the setup of the detectors is
asymmetric, and the signal of the leftmost sensor tends to reduce
the value of $x_{g}$. When $\varepsilon$ exceeds $\tau/2$, the
leftmost sensor is suppressed and sensor is added on the right
whose signal tends to increase the value of $x_{g}$ thereby
generating a discontinuity. The presence of the discontinuity is
signaled in the probability distribution of $x_{g}$ by a reduction
in the range of allowed values ($|x_{g}|<\tau/2$). With proper
selection of the two sampling functions H$(\omega)$, Eq. (20)
produces the discontinuity. Figure 4) plots  $x_{g}$ for two and
three sensors of relative dimensions sufficient to generate the
discontinuities. Discontinuities are present even in the $x_{g}$
calculated with the four, or more, sensors if the size of f$(x)$
is larger than the range covered by the set of detectors.
Reference~\cite{10} reports discontinuities for various sets of sensors
used by the authors. In all these cases, no polynomial
approximation of $x_{g}(\varepsilon)$ can cure the systematic
error. These singularities, which for even detector numbers are at
$\varepsilon=0$ and for odd detector numbers are at
$\varepsilon=\tau/2$, render very dangerous the reduction of the
noise fluctuation with the subtraction of a "bias". This operation
creates an admixture of different algorithms, (e.g.,  two and
three sensors) for $x_{g}$ with very different systematic errors
that are almost impossible to correct.

\subsection{Probability distribution of $x_{g}$}\indent

The above analysis and the explicit calculation of
$x_{g}(\varepsilon)$, enables calculating the probability of
having a value of $x_{g}(\varepsilon)$. We can compare this
probability with experimental histograms obviously keeping in mind
the differences discussed in Section 4. For a generic probability
distribution P$(\varepsilon)$ of $\varepsilon$ the probability of
$x_{g}(\varepsilon)$ is:
\begin{equation}
\Gamma(x_{g})=\mathrm{P}(\varepsilon)\Big|\frac{d\varepsilon}{dx_{g}}\Big|.
\end{equation}
For uniform distribution P$(\varepsilon)$, the probability
$\Gamma(x_{g})$ is proportional to the absolute value of the
derivative of $\varepsilon$  with respect to $x_{g}$. If the sign
of the derivative is the same (nonnegative) in the range of
$x_{g}$ values, the absolute value is useless. Now,
$x_{g}(\varepsilon)$ can be extracted from $\Gamma(x_{g})$, as
done in~\cite{16} for the $\eta$-function with a sample of equivalent
detectors. We will prove that the sign of $d\varepsilon/dx_{g}$ is
non-negative.

With our definitions of $\varphi(x)$, positive, symmetric, and
with a single maximum and the property of f$(x)$ to be positive
and maintaining a single maximum, $d\varepsilon/dx_{g}$ is always
non-negative. Here, we limit ourselves to f$(x)$ and $\varphi(x)$
functions that are continuous,  derivable, and have a single
maximum and $D>\tau$. More general functions (e.g. rectangular or
Dirac $\delta$-functions) will be late explored  with the
$D<\tau$.

The sign of $d\varepsilon/dx_{g}$  can be determined from the sign
of $dx_{g}/d\varepsilon$ which is more accessible. Equations (10)
and (13) are not suited to this task, because, being prepared for
simulations, their derivative can entail some complication. A
functional dependence of $x_{g}$ from $\varepsilon$, better suited
to derivatives, can be extracted from Eq. (3) applied to
S$_{\varepsilon}(\omega)$. Here, for uniform crosstalk and without
the application of the Poisson identity, S$_{\varepsilon}(\omega)$
is:
\[
\mathrm{S}_{\varepsilon}(\omega)=\sum_{n=-\infty}^{+\infty}\mathrm{f}(n\tau-\varepsilon)\mathrm{e}^{-in\tau\omega}
\]
and $x_{g}$ has its standard form:
\[
x_{g}(\varepsilon)=\sum_{n=-\infty}^{+\infty}n\tau\mathrm{f}(n\tau-\varepsilon).
\]
The derivative respect to $\varepsilon$ is:
\[
\frac{dx_{g}(\varepsilon)}{d\varepsilon}=\sum_{n=-\infty}^{+\infty}(-n\tau)\frac{d\mathrm{f}(y)}{dy}\Big|_{y=n\tau-\varepsilon}
\]
and, for our assumptions regarding the properties of f$(x)$,
$dx_{g}/d\varepsilon$ is positive $(D>\tau)$. For $n<0$
$d\mathrm{f}(y)/dy$ is calculated at $y=n\tau-\varepsilon$, i.e.,
before the maximum, it is positive, giving a positive contribution
to the sum. For $n>0$, $d\mathrm{f}(y)/dy$ is calculated at
$y=n\tau-\varepsilon$, i.e., after the maximum, it is negative
giving a positive contribution to the sum. For $D\leq\tau$, we
have $dx_{g}/d\varepsilon=0$ for $\varepsilon=0$.

In the case of the two sensors' COG and for the range $D$ of
$\varphi(x)$ greater than $\tau$, the derivative of $x_{g}$ with
respect to $\varepsilon$ must be calculated for $\varepsilon>0$
and $\varepsilon<0$ separately. The derivative for $\varepsilon>0$
is given by:
\[
\frac{dx_{g}}{d\varepsilon}=\frac{-\tau(1-x_{g}(\varepsilon)/\tau)}{\mathrm{f}(-\varepsilon)+\mathrm{f}(\tau-
\varepsilon)}\frac{d\mathrm{f}(y)}{dy}\Big|_{y=\tau-\varepsilon}+\frac{\tau\mathrm{f}(\tau-\varepsilon)}
{[\mathrm{f}(-\varepsilon)+\mathrm{f}(\tau-\varepsilon)]^{2}}\frac{d\mathrm{f}(y)}{dy}
\Big|_{y=-\varepsilon}\quad\varepsilon>0
\]
It is positive because $d\mathrm{f}(y)/dy$ at $y=\tau-\varepsilon$
is negative having been calculated after the maximum, $x_{g}/\tau$
is less or equal to 1/2, and $d\mathrm{f}(y)/dy$ at
$y=-\varepsilon$ is positive having been calculated before the
maximum. For $\varepsilon<0$ we get:
\[
\frac{dx_{g}}{d\varepsilon}=\frac{\tau(1+x_{g}(\varepsilon)/\tau)}{\mathrm{f}(-\varepsilon)+\mathrm{f}(-\tau-
\varepsilon)}\frac{d\mathrm{f}(y)}{dy}\Big|_{y=-\tau-\varepsilon}+\frac{-\tau\mathrm{f}(-\tau-\varepsilon)}
{[\mathrm{f}(-\varepsilon)+\mathrm{f}(-\tau-\varepsilon)]^{2}}\frac{d\mathrm{f}(y)}{dy}
\Big|_{y=-\varepsilon}\quad\varepsilon<0
\]
It is positive because $d\mathrm{f}(y)/dy$ at
$y=-\tau-\varepsilon$ is positive having been calculated before
the maximum, $x_{g}/\tau$ is greater or equal to -1/2, and
$d\mathrm{f}(y)/dy$ at $y=-\varepsilon$ is negative having been
calculated after the maximum. So, excluding  point $\varepsilon=0$
where $x_{g}(\varepsilon)$ has a discontinuity, its derivative is
positive and even the derivative of $\varepsilon$ with respect to
$x_{g}$ is positive.
Likewise, the derivative of
$x_{g}(\varepsilon)$ with respect to $\varepsilon$, for the three
sensors' COG is positive, here the discontinuities, if any, are at
the edge of the $\varepsilon$-distribution, and do not create
complications.

The distribution of probabilities, for two and three sensors is
illustrated in Figure 5. The signal distribution used is the same
used in  Figure 4.

\begin{figure}[h!]
\begin{center}
\includegraphics[scale=0.8]{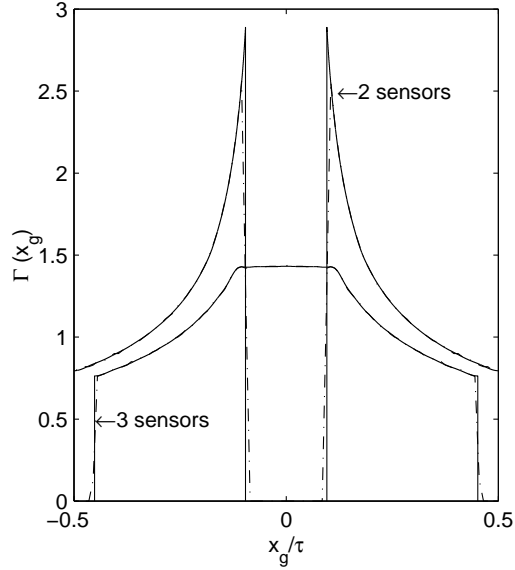}
\end{center}
\caption{\em Probability distributions of
$x_g$  for the two- and three-sensors COG. Solid lines are the
analytical calculations. Dot-dash lines join the
midsection of a scaled histogram bins generated on a sample of
$x_g$-data (bin size $x_g$/100). The differences are
concentrated in the curves' fast variation regions. The
discontinuities of Figure 3  are forbidden
$x_g$-values. }
\end{figure}

The dot-dashed lines indicate the histograms
normalized to the number of points and bin size. These plots
reveal the smearing effect in the regions of fast variation
$\Gamma(x_{g})$. This effect must be kept in mind when curves
extracted from the above equations are compared with experimental
distributions. A dramatic difference could be obtained when
$D=\tau$ and $\varphi(\pm\tau/2)=0$, here $\Gamma(0)=+\infty$, no
histogram can climb so high.

With our proof of non-negativity of the derivative, Eq. (24) can
be used as a differential equation for function
$\varepsilon(x_{g})$. The discontinuities of $x_{g}(\varepsilon)$
are no problem for the probability distribution $\Gamma(x_{g})$
because they imply forbidden values for $x_{g}$ with zero
probabilities, and, given the probability distribution of $x_{g}$
for a uniform $\varepsilon$ distribution, $\varepsilon(x_{g})$ is
given by:
\begin{equation}
\varepsilon(x_{g})=\tau\int_{-\tau/2}^{x_{g}}\Gamma(y)dy-\frac{\tau}{2}.
\end{equation}

This generalizes the method introduced in ~\cite{16} for the
$\eta$-function. The extraction of $\varepsilon(x_{g})$ from the
experimental data looks very easy.  All we have to do is integrate
a histogram of the frequencies of $x_{g}$ for a sample of the
equivalent signal distributions and detectors. In using Eq. (25),
we must keep in mind two warnings: 1) the fact that the histogram
is related to its generating function  in a complex way and 2) the
fact that noise drastically modifies Eq. (25):
\begin{equation}
\Gamma(x_{g})=\int d\xi_{1},\dots
d\xi_{n}\mathrm{P}(\varepsilon(x_{g},\xi_{1},\dots,\xi_{n}))
\mathrm{P}(\xi_{1})\dots\mathrm{P}(\xi_{n})\Big|\frac{\partial\varepsilon(x,\xi_{1},\dots,\xi_{n})}
{\partial x}\Big|_{x=x_{g}}.
\end{equation}
Now, the extraction of function $\varepsilon(x_{g})$ from this
relation is rather complicated, and we have to solve an integral
equation. The set of variables $\{\xi_{n}\}$ are the independent
sources of noise in the system. For example, in a two-strip COG,
three variables are easily encountered. Two  variables define the
electronic noises of the two strips with a  presumably Gaussian
distribution. The third variable is the total charge collected by
the strips, which has a Landau distribution. For a small amount of
noise, Eq. (26) can be solved as Eq. (25) yielding an acceptable
approximation of $\varepsilon(x_{g})$. The extraction of
$\varepsilon(x_{g})$ from the probability distribution of $x_{g}$
can be fruitfully used to correct the COG sampling error. The
presence of noise generally suggests the use of more than a
method. The two-sensor COG tends to be less sensitive around
$\varepsilon=0$, while conversely, the three-sensor COG is exact
for $\varepsilon=0$.  Careful simulations should determine the
optimal assembly of the two approaches.

Other information can be extracted from $dx_{g}/d\varepsilon$ in
the case of infinite sampling, and the absence of crosstalk. Here
Eq. (6) can be used, and the functions $x_{g}(\varepsilon)$ and
$\varepsilon(x_{g})$ are continuous and derivable
($dx_{g}/d\varepsilon$ is ($\Gamma(x_{g})^{-1}$). Deriving Eq. (6)
with respect to $\varepsilon$ gives:
\[
\frac{dx_{g}}{d\varepsilon}=1+2\sum_{k=1}^{+\infty}(-1)^{k}\cos
\big(\frac{2k\pi\varepsilon}{\tau}\big)\Phi(\frac{2k\pi}{\tau})
\]
which can be reassembled into:
\begin{equation}
\frac{dx_{g}}{d\varepsilon}=\sum_{k=-\infty}^{+\infty}\exp\big[i\frac{2k\pi}{\tau}
(\epsilon-\frac{\tau}{2})\big]\
\Phi(\frac{2k\pi}{\tau})=\tau\sum_{L=-\infty}^{+\infty}\varphi(\varepsilon-\frac{\tau}{2}-L\tau).
\end{equation}
The last expression, the Poisson identity, gives the periodic
extension of $\varphi(x-\tau/2)$, i.e., sum copies of
$\varphi(x-\tau/2)$ shifted by $L\tau$. These equations generalize
the results of~\cite{16} for any $D$. Now, if the support $D$ of
$\varphi(x)$ is less than or equal to $\tau$, the function
$\varphi(x)$ is reproduced in the interval $0<x<\tau$. Condition
$D\leq\tau$ is signaled by the absence of gaps in $\Gamma(x_{g})$,
and an infinite peak at $x_{g}=0$. The peak is given by the zeros
of $\varphi(x)$ at $|x|\leq\tau/2$. This pathology cannot be
reproduced by any histogram, and the reconstruction of
$\varphi(x)$ from experimental distributions unfortunately differ
from zero everywhere.

If $D$ is greater than $\tau$, tails of different periods add up
and other methods must be used to extract the form of
$\varphi(x)$: 1) a couple of sensors are considered together to
have an effective $D$ greater than $\tau$~\cite{16}, but if crosstalk
is present, it can survive the coupling. 2) $x_{g}(\varepsilon)$
is fit with Eq. (13) to get the form factors $\Phi(\gamma_{n})$
and reconstruct $\varphi(x)$ as given in Eq. (9), but no unique
results can be expected. 3) $\varepsilon(x_{g})$ is used to
reconstruct f$(x)$, and, going backward, to $\varphi(x)$ from Eq.
(10) in the absence of crosstalk. Any crosstalk present must be
known and Eq. (21) allows the extraction of $\varphi(x)$.

\subsection{Other properties of uniform crosstalk: the ideal detector}\indent

Equation (27) can be used to fix another  property of  functions
$\varphi(x)$ which are free of the COG discretization error. If
$x_{g}=\varepsilon$, than $dx_{g}/d\varepsilon=1$ and we have
$\tau\Sigma_{n}(x-n\tau)=1$. Here $D$ is always greater than
$\tau$. Functions of this type are introduced in~\cite{9} to give
consistent meaning to periodic functionals such as that used in
Eq. (3). These functions are called unitary functions of range
$\tau$. With the unitary functions, it is easy to verify the
absence of discretization error for a candidate function.  The sum
of $\varphi(x)$ with a set of shifted copy must give a unitary
function of range $\tau$. This can be verified by a graphic
method.

As we mentioned in  Section 5.2, the uniform crosstalk functions
p$(x,\tau_{1})$ have an identical property (aside from a
normalization factor). The conditions for uniform crosstalk are
identical to the absence of discretization error which, for
uniform crosstalk, must be  $\Sigma_{n}$p$(x-n\tau,\tau_{1})=1$.
The use of this relation is easier than the FT, and a great number
of uniform crosstalk functions can be generated and verified with
a simple graphic test.

Let us return to the set of uniform crosstalk functions  which are
free of  discretization error for any signal distribution. The
condition for the absence of discretization error can be
formulated as a condition on the extended periodic function. Now,
p$(x,\tau_{1})$ is uniform crosstalk and
$\Sigma_{n}$p$(x-n\tau,\tau_{1})=1$. The derivative
$d$P$(\omega,\tau_{1})/d\omega$ must be zero for
$\omega=2k\pi/\tau$ $(k=\pm 1,\pm 2,\pm 3,\dots)$ and for
$\omega=0$ due to the symmetry of P$(\omega,\tau_{1})$ and
p$(x,\tau_{1})$. A series similar to Eq. (27) constructed with
this $d$P$(\omega,\tau_{1})/d\omega$  has all the terms equal to
zero, and the extended periodic function
$\Sigma_{n}(x-n\tau)p(x-n\tau,\tau_{1})$ is also zero.

We can explicitly verify the absence of  discretization error
using a procedure like that used for  uniform crosstalk, but now
the sum of the shifted values of $x$p$(x,\tau_{1})$ must give  a
null function.  Starting from Eq. (24) and Eq. (14) for f$(x)$, we
can write  $x_{g}(\varepsilon)$ as:
\[
x_{g}(\varepsilon)=\sum_{n=-\infty}^{+\infty}n\tau\int_{-\infty}^{+\infty}
\mathrm{p}(n\tau-\varepsilon-x',\tau_{1})\ \varphi(x')dx'.
\]
Interchanging the sum and the integral, adding and subtracting
$(\varepsilon+x')$ to $n\tau$, and using the uniformity of
p$(x,\tau_{1})$, the normalization and symmetry of $\varphi(x)$,
we get:
\[
x_{g}(\varepsilon)=\varepsilon+\int_{-\infty}^{+\infty}\sum_{n=-\infty}^{+\infty}(n\tau-\varepsilon-x')
\mathrm{p}(n\tau-\varepsilon-x',\tau_{1})\ \varphi(x')dx'
\]
It is now evident that for uniform crosstalk functions with
$\Sigma_{n}(x-n\tau)$p$(x-n\tau,\tau_{1})=0$, the discretization
error disappears for any signal distribution or can be drastically
reduced as p$(x,\tau_{1})$ approximates this condition. It is easy
to verify that a triangular function of range $2\tau$ satisfies
the condition  $\Sigma_{n}(x-n\tau)$p$(x-n\tau,\tau_{1})=0$.

A detector with $\Sigma_{n}(x-n\tau)$p$(x-n\tau,\tau_{1})=0$ will be
defined {\em ideal detector}; it has $x_g(\varepsilon)=\varepsilon$
for any signal distribution.

\section{Noise, fluctuations, and border effects}
\subsection{Noise, cracks, and border effects}\indent

The various forms used above for the sampling function allow
exploring the efficiency modification and position reconstruction
in the presence of several different combinations of experimental
setups. Defining h$(x)$ by:
\[
\mathrm{h}(x)=a_{-1}\delta(x+\tau+\Delta_{-1})+a_{0}\delta(x-\Delta_{0})+
a_{1}\delta(x-\tau-\Delta_{1}),
\]
we can simulate the effects of the interstrip calibration errors
and the effects of cracks or incorrect detector positions.
Properly selecting  $\Delta_{-1}$, $\Delta_{0}$, and $\Delta_{1}$,
we get H$(\omega)$:
\[
\mathrm{H}(\omega)=a_{-1}e^{i\omega(\tau+\Delta_{-1})}+a_{0}e^{-i\omega\Delta_{0}}+
a_{1}e^{-i\omega(\tau+\Delta_{1})},
\]
H$(\omega)=$H$^{*}(-\omega)$ and H$(\omega)$ is no longer
symmetric in $\omega=0$. Equations (23) and (23') for
S$_{\varepsilon}$ and $x_{g}$ must be modified according to
requirements. In this case, the absence of symmetry in the
detector array gives a COG systematic error for $\varepsilon=0$
for three detectors and for $\varepsilon=\pm\tau/2$ for two
detectors.

Equation (21) enables modifying the response of even a  single
detector given by the function p$(x,\tau_{1})$, for instance, to
evaluate the effects of  different quality and size in a row of
crystals. In such case, different p$(x,\tau_{1})$ must be
considered, and different functions f$(x)$ must be sampled in the
proper positions. S$_{\varepsilon}(\omega)$ becomes:
\[
\mathrm{S}_{\varepsilon}(\omega)=\sum_{k=-M_{1}}^{M_{2}}\sum_{n=-\infty}^{+\infty}\Phi_{n}^{p}\
\exp(\frac{-i2n\pi\varepsilon}{T})\
\mathrm{P}_{k}(\frac{2n\pi}{T},\tau_{1})\mathrm{H}_{k}(\omega-\frac{2n\pi}{T})
\]

where H$_{k}(\omega)=a_{k}\exp[-i\omega(\tau_{k}+ k)]$; $x_{g}$ is
given by:
\[
x_{g}=\frac{1}{\mathrm{S}_{\varepsilon}(0)}\sum_{n=-\infty}^{+\infty}\Phi_{n}^{p}\
\exp(\frac{-i2n\pi\varepsilon}{T})\sum_{k=-\infty}^{+\infty}\mathrm{P}_{k}(\frac{2n\pi}{T})a_{k}
(\tau_{k}+\Delta_{k})\exp(\frac{i2n\pi(\tau_{k}+\Delta_{k})}{T}).
\]

Another fundamental effect to account for in the calculation of
the COG is noise.  There are essentially two main sources of
noise:  additive noise due to the readout electronics which
modifies the signal collected by each detector, and the
fluctuation of $\varphi(x)$ around its average.  Let us first
consider the readout electronic noise represented as an additive
noise. Now, the signals collected by the set of detectors are:
\[
\mathrm{sn}_{\varepsilon}(x)=\mathrm{f}^{p}(x-\varepsilon)\big[\sum_{m=-M_{1}}^{M_{2}}
\delta(x-m\tau)\big]+\sum_{m=-M_{1}}^{M_{2}}n_{m}\delta(x-m\tau),
\]
where $n_{m}$ are samples taken from the noise distribution scaled
by the noise-to-signal ratio, or better by  samples of its
distribution. This ratio is often far from being a constant. Now,
our equations represent a set of isolated points, each defined by
its corresponding noise samples. The expressions of the noisy COG
$nx_{g}(\varepsilon)$ and the total collected  signal
NS$_{\varepsilon}(0)$ are:
\begin{align*}
&\mathrm{NS}_{\varepsilon}(0)=\mathrm{S}_{\varepsilon}(0)+\sum_{m=-M_{1}}^{M_{2}}n_{m},\\
&nx_{g}(\varepsilon)=[x_{g}(\varepsilon)\mathrm{S}_{\varepsilon}(0)+\sum_{m=-M_{1}}^{M_{2}}
n_{m}m\tau]/\mathrm{NS}_{\varepsilon}(0),
\end{align*}
where $x_{g}(\varepsilon)$ and S$_{\varepsilon}(0)$ are functions
considered above in absence of noise.
\subsection{Fluctuations}\indent

Fluctuation of the signal distribution entails some complications.
Generally speaking, we can assume that $D$ is kept fixed, e.g., a
little larger than required, and the signal distribution
fluctuates as a random function of support $D$. In the case,
$\varphi_{r}(x)$ samples of $\varphi(x)$ can be generated with a
realistic distribution of the parameters defining $\varphi(x)$. As
in Eq. (9), we can put:
\[
\varphi_{r}(x)=\frac{1}{D}\Pi(\frac{x}{D})\big[\sum_{l=-\infty}^{+\infty}\Gamma_{l}\
\exp(i\frac{2l\pi}{D}x)\big],
\]
and, for the WKS theorem, the FT becomes:
\[
\Phi_{r}(\omega)=\sum_{l=-\infty}^{+\infty}\Gamma_{l}\
\frac{\sin(\omega D/2-l\pi)}{(\omega D/2-l\pi)}
\]
with $\Gamma_{l}^{*}=\Gamma_{-l}$ due to the reality of
$\varphi_{r}(x)$. The values of $\Phi_{r}(2n\pi/T)$ are the terms
to be inserted in  Eq.(21) to calculate $x_{g}$ and
S$_{\varepsilon}(0)$. The generation of $\Gamma_{l}$ samples could
be a complex operation. Theoretically, the procedure could be to
generate $\varphi_{r}(x)$, and than extract the parameters
$\Gamma_{l}$ with a FT. By our definitions, the average of all
samples $\varphi_{r}(x)$ is $\varphi(x)$.

A possible generation of  samples $\varphi_{r}(x)$ for
electromagnetic showers can be extraction  from the Monte Carlo
simulation. We consider an electromagnetic shower developing in a
homogeneous medium. The energy released is proportional to the
length of the paths of the conversion electrons and positrons. The
paths will be approximated to segments of a straight line in the
absence of a magnetic bending. The energy released by each segment
between two near parallel planes is proportional to the length of
the segment trapped between the two planes, and is zero if no
segment part is contained. A detailed demonstration is reported in
the next section. Observing that the total energy contribution
given by a segment is proportional to its length, we find the form
of $\varphi_{r}(x)$ given by a set of segments of length $L_{j}$
along $x$, starting from $x_{j}$, and with a total energy
contribution $s_{j}$:
\begin{equation}
\varphi_{r}(x)=\frac{1}{\sum_{j}s_{j}}\big[\sum_{j}\frac{s_{j}}{L_{j}}\
\Pi(\frac{x-x_{j}-L_{j}/2} {L_{j}})\big].
\end{equation}
Its FT $\Phi_{r}(\omega)$ is:
\begin{equation}
\Phi_{r}(\omega)=\frac{1}{\sum_{j}s_{j}}\Big\{\sum_{j}s_{j}\
\exp[-i\omega(x_{j}+L_{j}/2)]\ \frac{\sin(\omega L_{j}/2)}{\omega
L_{j}/2}\Big\}.
\end{equation}

The construction of $\varphi_{r}(x)$ and $\Phi_{r}(\omega)$ is now
easer. We need to extract a set of values $\{s_{j},x_{j},L_{j}\}$
for each Monte Carlo event, and, accumulating a sufficient number
of events, we can generate the average distribution $\omega(x)$
with a realistic sample of fluctuations. For our needs, not all
the details of Equations (28) and (29) are worthy of attention.
The convolution of $\varphi_{r}(x)$  with the p$(x,\tau_{1})$ is a
low-pass filter which considerably smears $\varphi_{r}(x)$. In our
approach based on the FT and FS, this means that the sum over a
discrete index can be safely cut off at some reasonable value
without negatively effecting the results. For this reason, even a
statistical generation of the parameters $\{s_{j},x_{j},L_{j}\}$
rougher than a detailed Monte Carlo calculation should suffice for
fine-tuning a position reconstruction algorithm.

Lastly, let us examine the asymmetry introduced by the slight
($3^{0}$) off-axis distortion of the crystals of the CMS
em-calorimeter. In this case, the average shower is not symmetric:
$x_{g}$ and $\varepsilon$ now refer to the shower's COG, which,
when projected on the $x$-axis, does not coincide with the impact
point of the photon (or electron/positron) on the calorimeter
face. To extract the impact point, we must first correct the
discretization on $\varepsilon$. With $\varepsilon$ and the
average depth of the shower's COG, we can calculate the shower's
impact point.
\subsection{Signal density for a set of tracks}\indent

Let us consider the energy released along the path of a charged
particle, assuming that the signal collected is proportional to
the length of the path. Neglecting its width, we can represent a
charged track   as a set of segments of a straight line whose
length is $s_{j}$, whose starting point is
$\mathbf{r}_{j}\equiv\{x_{j},y_{j},z_{j}\}$ and whose projections
on the axis are $\mathbf{L}_{j}\equiv\{X_{j},Y_{j},Z_{j}\}$. The
three-dimensional energy distribution of each segment can be
written as:
\begin{equation}
\mathrm{e}_{j}(x,y,z)=s_{j}\int_{0}^{1}d\lambda\delta(x-x_{j}-\lambda
 X_{j})\delta(y-y_{j}-\lambda Y_{j})\delta(z-z_{j}-\lambda Z_{j}).
\end{equation}

The energy density collected at point $x$ is given by:
\[
\mathrm{e}_{j}(x)=\int_{R^{2}}dydz\mathrm{e}_{j}(x,y,z).
\]
Since the function e$_{j}(x,y,z)$ is too singular to be handled,
it is better to use its FT, (E$_{j}(\mathbf{W})$ is a compact
notation for E$_{j}(\omega_{x},\omega_{y},\omega_{z})$).
\[
\mathrm{E}_{j}(\mathbf{W})=\int_{R^{3}}d^{3}r\
\mathrm{e}_{j}(\mathbf{r})\ \exp[-i\mathbf{W\cdot r}]
\]
which, substituting Eq. (30), becomes:
\[
\mathrm{E}_{j}(\mathbf{W})=s_{j}\int_{0}^{1}d\lambda\
\exp[-i\mathbf{W}\cdot(\mathbf{r}_{j}+\lambda\mathbf{L}_{j})]=
\exp[-i\mathbf{W}\cdot(\mathbf{r}_{j}+\mathbf{L}_{j}/2)]\
\frac{\sin(\mathbf{W\cdot L}_{j}/2)}{\mathbf{W\cdot
L}_{j}/2}s_{j}.
\]
The energy distribution e$_{j}(x)$ is given by the inverse FT of
E$_{j}(\omega_{x},0,0)$:
\[
\mathrm{E}_{j}(\omega_{x},0,0)=s_{j}\
\exp[-i\omega_{x}(x_{j}+X_{j}/2)]\frac{\sin(\omega_{x}X_{j}/2)}{(\omega_{x}X_{j}/2)}.
\]
It is evident that E$_{j}(\omega_{x},0,0)$ is the FT of:
\[
\varphi_{j}(x)=\frac{s_{j}}{X_{j}}\
\Pi(\frac{x-x_{j}-X_{j}/2}{X_{j}}).
\]
Summing over the $j$-index and normalizing, we get Eq. (28). The
periodic extension of $\varphi_{j}^{p}(x)$ can be achieved
according to the rule:
\[
\varphi_{r}^{p}(x)=\frac{1}{\sum_{j}s_{j}}\sum_{j}\varphi_{j}^{p}(x)=\frac{1}{T\sum_{j}s_{j}}
\sum_{n=-\infty}^{+\infty}\sum_{j}\mathrm{E}_{j}(\frac{2n\pi}{T},0,0)
\mathrm{e}^{-i2n\pi x/T}
\]
Bidimensional distributions can be generated in a similar way. At
a reasonable cutoff ($n\approx 200-250$), the interferences among
the contributions of different segments smear the corners and
suppress the Gibbs effects at each rectangle's borders in
$\varphi_{j}^{p}(x)$ giving very realistic energy distributions.
This form plots easier than Eq.(26).   To see the aspect of
$\varphi_{j}^{p}(x)$, we generate a few random distributions of
segments whose lengths follow a Rayleigh distribution averaging
around one radiation length.

\begin{figure}[h!]
\begin{center}
\includegraphics[scale=0.8]{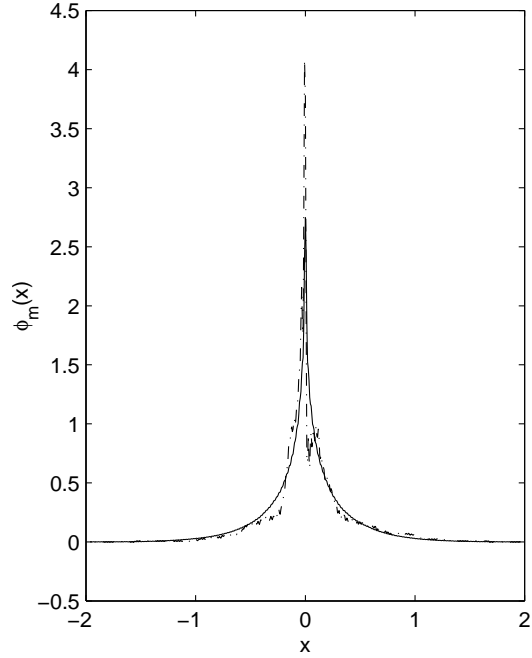}
\end{center}
\caption{\em Signal distribution of a set of random segments
generated as explained in the text. Dotted line indicates a
random sample; solid line is the average of 150
samples. The horizontal scale is in unit of  Molier radiuses. }
\end{figure}
\begin{figure}[h!]
\begin{center}
\includegraphics[scale=0.6]{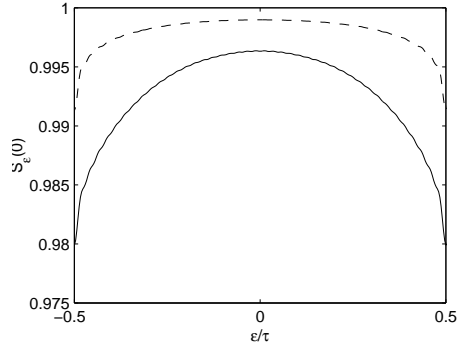}
\end{center}
\caption{\em Efficiency plots in function of $\varepsilon/\tau$
for three-sensor (solid line) and five-sensor (dashed line)
signal collection, with effective loss of 0.002 at the sensor
borders. The sensor size ($\tau$) is 1.1 Molier radius.}
\end{figure}

Their origins and zenith angles are
Gaussian functions with standard deviations respectively
($0.5,\pi/3$).
The azimuth angles have uniform distributions. $z$-distributions
of the segment origins are irrelevant to this simulation. The
average $x$-distribution of the sets of random segments can be
made to resemble to the average em shower described in~\cite{12}. One
of the characteristics they share is the presence of a narrow peak
at $x=0$ and a near exponential decrease in the shoulders (Figure
6) . We use this distribution, scaled with the Molier radius, to
simulate a more complex signal distribution than that in Eq.(8).
Due to some similarity with an em shower, we sampled this signal
with sensors of the size (in Molier  radiuses) used in the CMS em
calorimeter.

Figure 7) shows the efficiencies versus   of a set of
three and five sensors, with a total loss at the borders of 0.002.
Figure 8) shows the systematic error of the two- and three-sensor
COG for this signal distribution.
~\begin{figure}[h!]
\begin{center}
\includegraphics[scale=0.7]{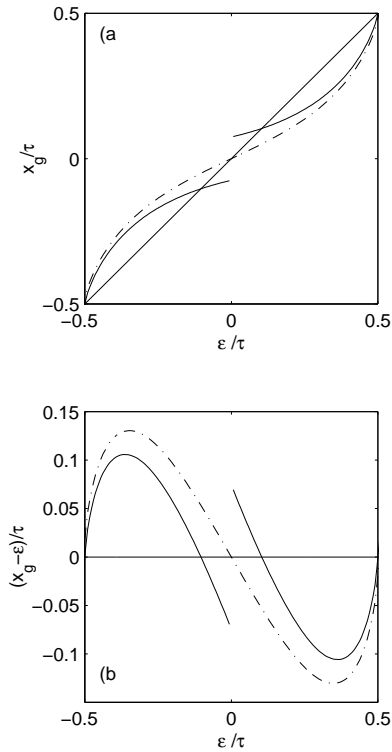}
\end{center}
\caption{\em The same plots as Figure 4 for the average signal
distribution of  Figure 6. }
\end{figure}

\section{Conclusions}\indent

We have analytically explored many properties of the center of
gravity (COG) algorithm. The effects of the sampling destroy the
simple identity of the COG with the symmetry axis of the signal
distribution. Few exceptions are worthy of mention.  The symmetric
functions with band limitation $\Phi(\omega)=0$ for
$\omega>2\pi/\tau$ have the sampled COG coincident with the
symmetry axis of $\varphi(x)$. This property is shared even by
rectangular functions with $D=\tau,2\tau,3\tau,\dots k\tau\dots$
multiples of the sampling interval, the triangular functions with
$D=2\tau,4\tau,6\tau\dots$ and all the convolutions of $n$
identical rectangular functions  for $D=n\tau,2n\tau,3n\tau\dots$.
In all other cases, the COG has a complex nonlinear relation with
the symmetry axis of $\varphi(x)$ which can be handled with data
involving $\varphi(x)$. An interesting effect is provided by the
crosstalk, which is normally considered a sort of undesired
distortion introduced in signal collection. It has been proven
that the crosstalk can be modulated to have a signal collection
whose COG turns out to be free of discretization error for
{\it{any signal distribution}}. One of the simplest crosstalk
forms with the above property is a triangle with size $2\tau$, but
the convolution of any symmetric function with a triangular shape
maintains the property. An easy test for the crosstalk function to
be of the type described is
$\sum_{n}(x-n\tau)$p$(x-n\tau,\tau_{1})=0$. If a large sample of
equivalent signal distribution collections is available, it can be
used to extract a faithful approximation to $\varepsilon(x_{g})$
from a distribution of $x_{g}$, and, if $D<\tau$, to extract the
signal distribution $\varphi(x)$ or the detector response
function. If precise position measurements are available, the
mismatch of the COG from can be used to access to $\varphi(x)$ or
to extract the detector response function. Explicit equations are
given for the simulation of the experimental setup, and for the
fine-tuning of the algorithm in the presence of discontinuities.
The relations of $x_{g}$ with $\varepsilon$ show that the COG,
calculated with the suppression of the low-signal detectors, must
be carefully handed to avoid the admixtures of different numbers
of sensors in the calculation, which, for the different systematic
errors, are difficult to correct. Various models of noise and
fluctuations are explored with an emphasis on the extraction of
signal distributions and averages from a Monte Carlo simulation.
Although the equations presented look very complex, they are
actually very easy to use. Only a few lines of MATLAB programming
language are needed to implement each one and to generate a
complete simulation of a measurement setup with noise and
fluctuations. All these results remain valid for asymmetric signal
distributions, with only negligible modifications to the
equations. In this case, $\varepsilon$ indicates the signal
distribution's COG and no longer its impact point which we assumed
to coincide with the symmetry axis.  The detection of the impact
point requires further {\it{ad hoc}} assumptions.




\end{document}